%% file: tester.tex
\documentclass[12pt]{article}        %%%%%%%%%%%%%%%%%%
\usepackage{pslatex}
\usepackage{amsmath}
\usepackage{amssymb}
\usepackage{euscript}
\usepackage{epsfig}

\usepackage{cite}

\usepackage{alltt}

\usepackage{epic}
\usepackage{color}
\usepackage{subfigure}
\newif\ifpdf
\ifx\pdfoutput\undefined
    \pdffalse           % we are not running PDFLaTeX
\else
    \pdftrue
\fi
\ifpdf
    \usepackage{pslatex}
\fi

\begin{document}                     %%%%%%%%%%%%%%%%%%%%%%

\allowdisplaybreaks

%%%%%%%%%%%%%%%%%%%%%%%%%%%%%%%%%%%%%%%%%%%%%%%%%%%%%%%%%%%%%%%%%%%%%%%%%%%%%%%%%
%%%%%%%%%%%%%%%%%%%%%%%%%%%%%%%%%%%%%%%%%%%%%%%%%%%%%%%%%%%%%%%%%%%%%%%%%%%%%%%%%
\begin{titlepage}

\begin{flushright}
%\date{today}
 {\bf CERN-TH/2002-271}\\ 
 {\bf  TP-U\'Sl-01/02}
\end{flushright}
\vspace{1mm}
\begin{center}{\bf\Large  MC-TESTER: a universal tool}\end{center}
\begin{center}{\bf\Large   for comparisons of Monte Carlo predictions}\end{center}
\begin{center}{\bf\Large for particle decays in high energy physics $~^{\dag}$ }\end{center}
\vspace{3 mm}
\begin{center}
{\large \bf P. Golonka$^{a,b}$,}
{\large \bf T. Pierzcha\l a$^{c}$,}
{\large \bf Z. W\c{a}s$^{d,e}$}\\
\vspace{1 cm}
{\em $^a$ Faculty of Nuclear Physics and Techniques, University of Mining and 
Metallurgy\\ Reymonta 19, 30-059 Cracow, Poland.  }\\
{\em $^b$CERN, EP/ATR Division, CH-1211 Geneva 23, Switzerland.}\\
{\em $^c$ Institute of Physics, University of Silesia,\\ Uniwersytecka 4, 
40-007 Katowice, Poland.}\\
{\em $^d$Institute of Nuclear Physics,
         Radzikowsiego 152 , 31-342 Cracow, Poland.}\\
{\em $^e$CERN, Theory Division, CH-1211 Geneva 23, Switzerland.}\\

\end{center}

\vspace{1mm}
\begin{abstract}
Theoretical predictions in high energy physics are routinely provided 
in the form of Monte Carlo generators.
Comparisons of predictions from different programs and/or different 
initialization set-ups are often necessary. 
  {\tt MC-TESTER} can be used
for such tests of decays of intermediate 
states (particles or resonances) in a semi-automated way.

Our test consists of two steps. Different Monte Carlo
programs are run; events with decays of a chosen 
particle are searched, decay trees are analyzed and appropriate information is stored. 
Then, at the analysis step, a list of all found decay modes is defined 
and branching ratios
are calculated for both runs.
Histograms of all scalar Lorentz-invariant masses constructed from the decay products are plotted
and compared for each decay mode found in both runs. For each plot
a measure of the difference of the distributions 
is calculated 
and its  maximal value over all  histograms 
for each  decay channel is printed in a summary table.
%Several algorithms for {\tt Shape Difference Parameter} calculation are provided; a user may
%provide custom algorithms as well.

As an example of
{\tt MC-TESTER} application,
we include a test with the $\tau$ lepton decay Monte Carlo generators,
{\tt TAUOLA} and {\tt PYTHIA}.
The {\tt HEPEVT} (or {\tt LUJETS}) common block is used as exclusive source  
of information on the generated events.
%In future, the program is expected to
%work with other generators and event data structures of FORTRAN and also of
%other languages such as {\tt C++}  or {\tt Java} as well.

\end{abstract}
\begin{center}
{\it Published in Computer Physics Communications, {\bf 157}(2004) 1}
\end{center}

\vspace{1mm}
\begin{flushleft}
{\bf CERN-TH/2002-271}\\
 {\bf  TP-U\'Sl-01/02} \\
{\bf version 1.1, December 2002}
\end{flushleft}

\vspace{1mm}
\footnoterule
\noindent
{\footnotesize
\begin{itemize}
\item[${\dag}$]
This work is partially supported by
the Polish State Committee for Scientific Research 
(KBN) grants 2 P03B 001 22, %(zbw),  
5P03B10721 %(tomek), % Tomka grant 
and also by the European Community's Human Potential
Programme under contract HPRN-CT-2000-00149 Physics at Colliders.

\end{itemize}
}

\end{titlepage}

\tableofcontents 

\newpage
%%%%%%%%%%%%%%%%%%%%%%%%%%%%%%%%%%%%%%%%%%%%%%%%%%%%%%%%%%%%%%%%%%%%%%%%%%%%%%%%%
%%%%%%%%%%%%%%%%%%%%%%%%%%%%%%%%%%%%%%%%%%%%%%%%%%%%%%%%%%%%%%%%%%%%%%%%%%%%%%%%%
%%%%%%%%%%%%%%%%%%%%%%%%%%%%%%%%%%%%%%%%%%%%%%%%%%%%%%%%%%%%%%%%%%%%%%%%%%%%%%%%%

\noindent{\bf PROGRAM SUMMARY}
\vspace{10pt}

\noindent{\sl Title of the program:} \- {\tt MC-TESTER}, version 1.1

\noindent{\sl Computer:}\-
PC, two Intel Xeon 2.0 GHz processors , 512MB RAM 

\noindent{\sl Operating system:}\-
Linux Red Hat 6.1, 7.2, and also 8.0

\noindent{\sl Programming language used:}\-
{\tt C++, FORTRAN77}: gcc 2.96 or 2.95.2 (also 3.2) compiler suite with {\tt g++} and {\tt g77}

\noindent{\sl Size of the package:}\\
 7.3 MB  directory including example programs
 (2 MB compressed distribution archive), without
{\tt ROOT} libraries (additional 43 MB).

\noindent{\sl Additional disk space required:}  \\
Depends on the analyzed particle:
 40 MB  in the case of $\tau$ lepton decays (30 decay channels, 594 histograms,
 82-pages booklet).


\noindent{\sl Nature of the physical problem:}\\
%---------------------------------------------
The decays of individual particles are  well defined
modules of a typical Monte Carlo program chain in high energy 
physics.
A fast, semi-automatic way of comparing results from different programs is 
often desirable, for the development of  new programs, to check 
correctness of the installations or for discussion of  uncertainties.

\noindent{\sl Method of solution:}\\
%---------------------------------
A typical HEP Monte Carlo program stores the generated events in
the event records such as {\tt HEPEVT} or  {\tt PYJETS}. {\tt MC-TESTER} scans, 
event by event, the contents of the record   and searches for the decays of 
the particle under study. The list of the found decay modes is successively 
incremented and histograms of  all invariant masses 
which can be calculated from the momenta of the particle decay products  
are defined and filled. The outputs from the two runs of distinct
programs can be later compared. A booklet of comparisons is 
created:  for every decay channel,
all histograms present in the two outputs are plotted and parameter
 quantifying shape difference 
is calculated. Its maximum over every decay channel is printed in the
 summary table. 

\noindent{\sl Restrictions on the complexity of the problem:}\-
%------------------------------------------------------------
For a list of limitations see section \ref{sec:outlook}. 

\noindent{\sl Typical running time:}\\
%-----------------------------------
Varies substantially with the analyzed decay particle.
On a PC/Linux with 2.0 GHz processors 
{\tt MC-TESTER} increases the run time of  the $\tau$-lepton Monte Carlo
program   {\tt TAUOLA} by $4.0$ seconds for every $100~000$ analyzed events
(generation itself takes $26$ seconds).
The analysis step takes $13$ seconds; \LaTeX ~processing takes additionally $10$ seconds.
Generation step runs may be executed simultaneously on multi-processor machines.

\noindent{\sl Accessibility: } \\
%-----------------------------------
web page: {\tt http://cern.ch/Piotr.Golonka/MC/MC-TESTER} \\
e-mails:\hspace*{4.5mm} {\tt Piotr.Golonka@CERN.CH},\\\hspace*{18.5mm} 
{\tt T.Pierzchala@friend.phys.us.edu.pl},
\\\hspace*{18.5mm} {\tt Zbigniew.Was@CERN.CH}.
\newpage
%%%%%%%%%%%%%%%%%%%%%%%%%%%%%%%%%%%%%%%%%%%%%%%%%%%%%%%%%%%%%%%%%%%%%%%%%%%%%%%%%%%%%%%%%%%%%
%%%%%%%%%%%%%%%%%%%%%%%%%%%%%%%%%%%%%%%%%%%%%%%%%%%%%%%%%%%%%%%%%%%%%%%%%%%%%%%%%%%%%%%%%%%%%

\section{Introduction}
%%%%%%%%%%%%%%%%%%%%%%%%%%%%%%%%%%%%%%%%%%%%%%%%%%%%%%%%%%%%%%%%%%%%%%%%%%%%%%%%%%%%%%%%%%%%%

In the phenomenology of high-energy physics, the question of establishing
uncertainties for theoretical predictions used in the interpretation 
of the experimental data is of high importance. For example,
at the time of the experiments at LEP
 several workshops (see \cite{MC2000}) were devoted to this question.
As the required precision is often high, theoretical predictions need 
to be presented in the form of Monte Carlo event generators; all detector 
effects can 
therefore be easily combined with the theoretical ones, using
rejection methods.

Whenever possible, theoretical predictions are separated into individual 
building blocks, which are later combined into complicated 
Monte Carlo generator systems for the complete predictions.
A good example of such a building block is the generation of a decay of a
particle such as the $\tau$ lepton.
While working on {\tt TAUOLA} \cite{Jadach:1993hs,Jezabek:1991qp,Jadach:1990mz}, 
we devised a set of tests that performed
comparisons of results produced by two versions of the program. 
We  realized that a test that compares
all distributions of invariant masses built from the decay products
of the $\tau$ lepton (in a particular decay channel) gives a valuable
and quite complete answer.

In the present paper we document a new tool, {\tt MC-TESTER}, which
 tests/compares in an automated way  particle decays
generated by two MC programs. The
analysis  consists of two steps.
First, appropriate data have to be collected from a run of every tested program. 
To this end  the libraries of {\tt MC-TESTER}  have to be loaded, 
a testing routine 
called, and an identifier of the particle to be analyzed has to be provided.
Event after event,
 {\tt MC-TESTER} will read data from consecutive
event records (such as {\tt HEPEVT}) and  look for decays 
of the tested particle. 
Once  found, a decay channel is identified from its
decay products. On the first occurrence of the particular 
decay channel, it is added to 
a list of found decay modes;   all histograms for invariant 
masses that can be formed of decay products are initialized and filled
for the first time. For
later occurrences, appropriate histograms are simply filled.
At the end of the run all information is stored to a file.

In the second step, the results obtained from the two generation runs from two
 Monte Carlo programs are compared and a booklet is created. It includes
a table of decay modes found in the two runs with corresponding branching ratios.
For matching decay channels (present in outputs from both generators), 
comparison plots are provided. For each invariant mass distribution,  histograms
from the two runs are plotted; their ratio (after normalization) is also plotted
in the same frame. The {\tt Shape Difference Parameter} ({\tt SDP})
 is calculated
and printed on the plot as well. The  maximum of {\tt SDP} over all plots 
for the given decay channel is printed in the table of decay modes as well.

The paper is organized as follows:
section \ref{sec:installation}  explains how the first step of the program (generation)
should be installed and executed. Section \ref{sec:analysis} explains the second 
step of the  {\tt MC-TESTER} action, namely the analysis; 
a description of the program output is also given in that section. 
Section \ref{sec:SDP-algorithms} includes a  description of the
available algorithms for {\tt SDP} calculation. 
Section \ref{sec:package-organization} is devoted to the description of the package,  
directory organization, and technical information on its use;
further details and explanation of input parameters may be found in 
the appendix.
Section \ref{sec:outlook} closes the documentation with a discussion of package limitations and 
possible future extensions.

%Appendix A (\ref{appendix.A}) includes descriptions of the input file 
%({\tt SETUP.C}) and {\tt FORTRAN} routines, which are used to control 
%the values of parameters used by {\tt MC-Tester}.

%%%%%%%%%%%%%%%%%%%%%%%%%%%%%%%%%%%%%
\section{Installation and generation step}
\label{sec:installation}

{\tt MC-TESTER} is distributed in a form of an archive containing source files.
Currently only Linux operating system is supported: other systems may be
supported in the future if sufficient interest is found. We have checked {\tt MC-TESTER}
on Red Hat 6.1 and Red Hat 7.2 installations.
In order to run {\tt MC-TESTER} one needs:
\begin{itemize}
    \item {\tt gcc} 2.96 or 2.95.2 compiler suite with {\tt g++} and {\tt g77}
     installed,
    \item {\tt ROOT} package properly installed and set up (please refer to 
        \cite{root-install-www}  or {\tt ROOT\_INSTALL} file in {\tt doc/} 
        subdirectory for details),
    \item \LaTeX\  package.
\end{itemize}
After unpacking, one needs to compile {\tt MC-TESTER} libraries using
{\tt make} command executed in its main directory. If completed successfully, 
the user is instructed how to proceed with
the example tests. Examples for the {\tt MC-TESTER} use based on 
the $\tau$ decay generators {\tt TAUOLA} and {\tt PYTHIA}
are distributed together with the package; they reside in the {\tt examples-F77/}
subdirectory.

{\tt MC-TESTER} distribution is
a complete, ready-to-use testing environment, with subdirectories
dedicated to generation and analysis steps (see  section \ref{dir-struct} for details),
and run-time parameters controlled by simple configuration files
({\tt SETUP.C} - see section \ref{SETUP.C} and the Appendix ).

The {\tt TAUOLA} test
 (located in {\tt examples-F77/tauola/})
is intended to show how {\tt MC-TESTER} can be used 
to compare two versions of the same generator. The versions (CLEO and ALEPH) 
have been prepared using {\tt TAUOLA-PHOTOS-F}
package \cite{Golonka:2002iu} and are placed in the {\tt gen1/} and {\tt gen2/} subdirectories.

In order to prepare this example,  {\tt make} in 
{\tt examples-F77/tauola/} directory should be executed. For two versions of {\tt TAUOLA}, 
demo programs linked with the {\tt MC-TESTER}'s 
libraries will be prepared, compiled and linked 
({\tt gen1/tautest.exe}  and {\tt gen2/tautest.exe}),
and  runs 
are  ready for execution with {\tt make run1} or {\tt make run2}, 
respectively. Alternatively, one may go to
the subdirectory {\tt gen1/} or {\tt gen2/} and execute {\tt ./tautest.exe} there.

The {\tt FORTRAN} demo programs 
( {\tt examples-F77/tauola/gen1/tautest.f} and 
\\ {\tt examples-F77/tauola/gen2/tautest.f} )
created from {\tt examples-F77/tauola/tautest.F} at the pre-compilation step,
initialize the {\tt TAUOLA} and {\tt MC-TESTER}. In particular, 
demos are set 
to generate $100~000$ events.  In the code,  one can easily find (see also sections 
\ref{other-gen:F77} and \ref{SETUP_in_F77})
  how the routines for 
{\tt MC-TESTER}'s operation: {\tt MCTEST(MODE), MCSETUP(WHAT,VALUE)} 
are called. At the end of generation,
 the {\tt MC-TESTER} finalization routine {\tt MCTEST(1)}   stores
the output file: {\tt mc-tester.root}.
The output data files 
({\tt gen1/mc-tester.root} and {\tt gen2/mc-tester.root}) 
are then ready to be transferred to the {\tt MC-TESTER}'s analysis part 
(see Section \ref{sec:analysis}) using e.g. {\tt make move1}, 
{\tt make move2} commands.

In order to perform a comparison test of $\tau$ decay in  {\tt PYTHIA},
one needs to compile and execute the code placed in {\tt examples-F77/pythia/}.
To compile the {\tt PYTHIA} library and the main example program 
(the source code in {\tt pythiatest.f}) 
one can execute the {\tt make} command. To  run it, the command
{\tt make run} can be executed (or directly {\tt pythiatest.exe}). 
By default $10~000$ events are set to be generated.
As in the case of the {\tt TAUOLA} example program, the output histograms are stored 
in the {\tt mc-tester.root} file and 
 may be copied to the analysis directory using
the {\tt make move1} or {\tt make move2} commands.

It is possible to control the {\tt MC-TESTER}'s parameters using the {\tt SETUP.C} file.
The syntax of the file is described briefly in section \ref{SETUP.C}, and in
the Appendix \ref{appendix.Setup-parameters}. The {\tt SETUP.C} file needs to 
be put in the directory from which
generation program is being executed (usually it is the same directory in which
binary executable file exist). Examples of {\tt SETUP.C} files are already present
in example generation directories: they are used to set the description of
the generator being run.

The output data file is usually put in the directory in which generation program was
executed. The name of the file and the path may however be changed using the {\tt SETUP.C}
file (see \ref{appendix.S.result1-path}, \ref{appendix.S.result2-path}).

The issue of the {\tt MC-TESTER} use with ``any'' Monte Carlo generators is addressed
in section \ref{other-generators}. We want to stress, that it is relatively easy to use {\tt MC-TESTER}
with  a Monte Carlo event generator, which the user wants to test: it is sufficient to 
link the {\tt MC-TESTER}'s libraries, the {\tt  ROOT} libraries and to insert 
three subroutine calls into 
the user's code: for the {\tt MC-TESTER} initialization, finalization and analysis.

For the users interested in trying only the analysis part of {\tt MC-TESTER} 
(Section \ref{sec:analysis}), and to avoid a lengthy generation phase,
a ready-to-use data files are provided in the directory 
{\tt examples-F77/pre-generated/}. There, the  {\tt MC-TESTER}'s 
{\tt mc-tester.root} files
(produced by very long runs with {\tt TAUOLA} and {\tt PYTHIA}), are stored. To copy the files to 
the directories of the analysis step, 
the command {\tt make move} can be used, similarly as explained above.

%%%%%%%%%%%%%%%%%%%%%%%%%%%%%%%%%%%%%%%%%%%%%%%%%%%
\section{Analysis}
\label{sec:analysis}
\begin{figure*}
\begin{center}
\ifpdf
    \subfigure[The first page of analysis booklet.]{\includegraphics[  angle=90, width=0.50\paperwidth, keepaspectratio]{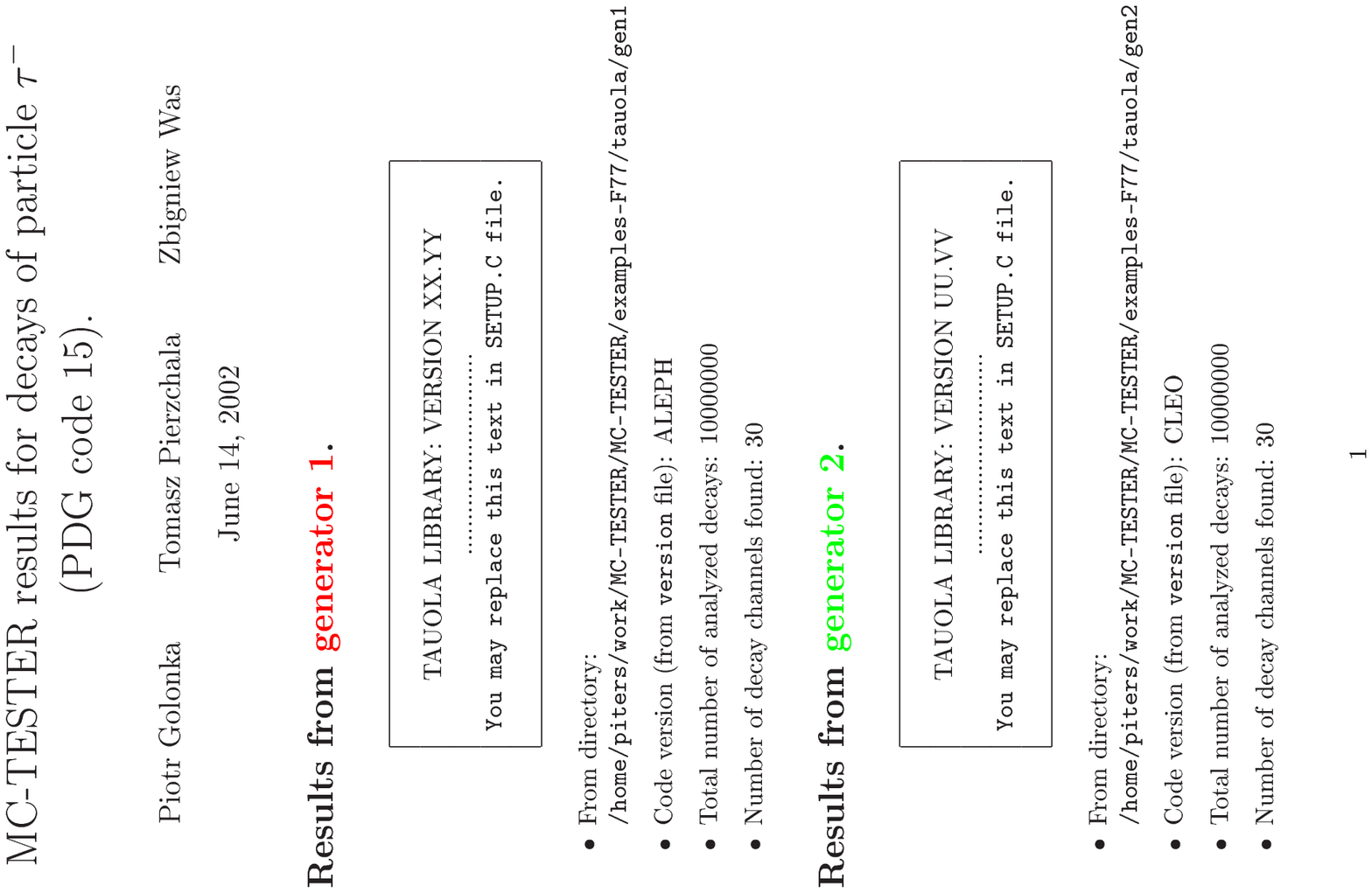}\label{subfig:firstpage}}\\
    \subfigure[The table of found decay channels.]{\includegraphics[  angle=90, width=0.50\paperwidth,keepaspectratio]{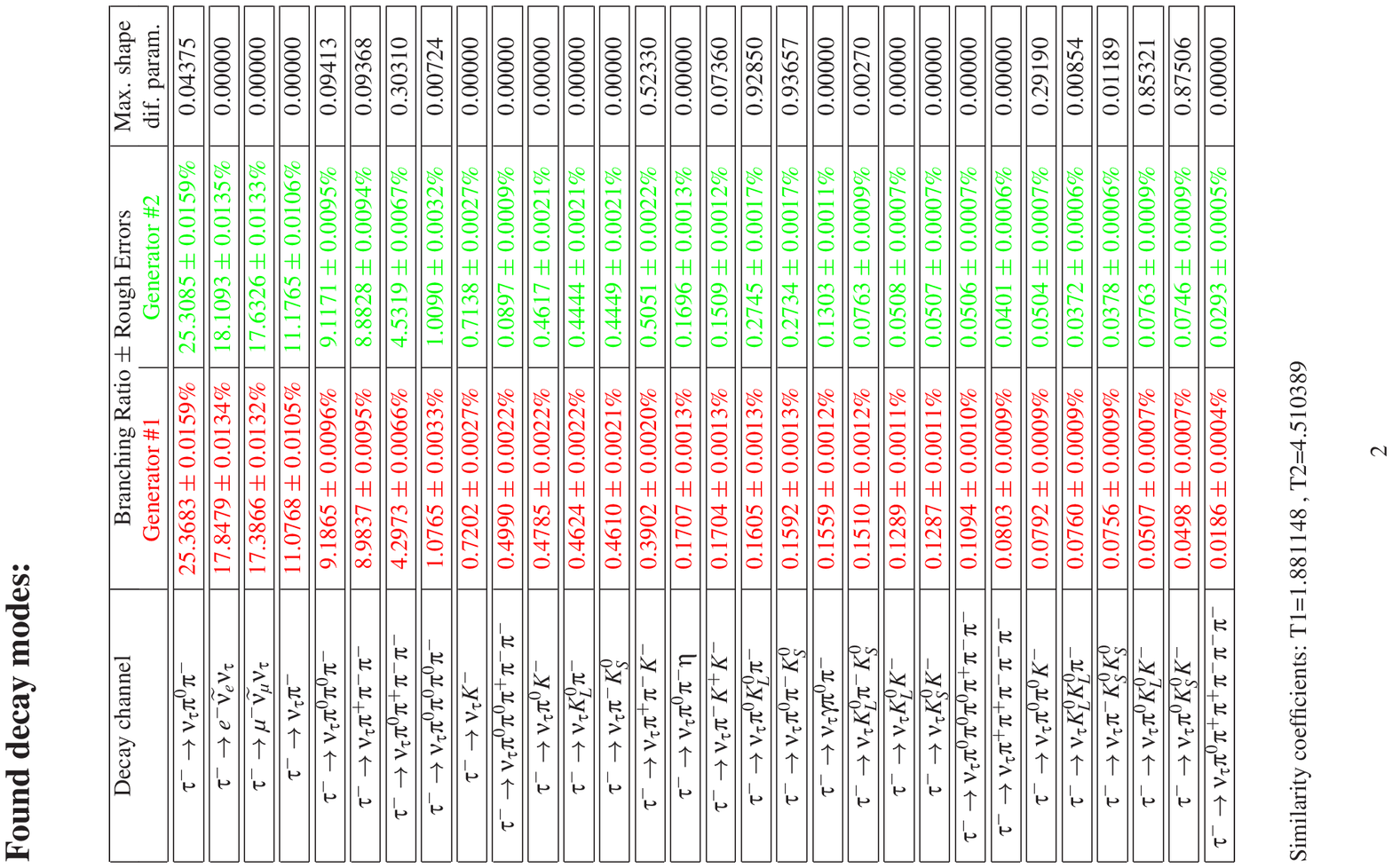}\label{subfig:channelstable}}
\else
    \subfigure[The first page of analysis booklet.]{\includegraphics[  bb=120 150bp 500bp 842bp,
      clip, angle=90, width=0.72\paperwidth, keepaspectratio]{tester1}\label{subfig:firstpage}}\\
    \subfigure[The table of found decay channels.]{\includegraphics[  bb=120 150bp 520bp 842bp,
      clip, angle=90, width=0.72\paperwidth,keepaspectratio]{tester2}\label{subfig:channelstable}}
\fi
    \caption{ Example of booklet's informational pages produced at analysis step.
    At the bottom of the table \ref{fig-booklet1}(b), the $T_1$, $T_2$ coefficients
    quantifying the difference in all decay channels combined are printed (see chapter
    \ref{subsec:recent_updates} for details). }
\label{fig-booklet1}
\end{center}
\end{figure*}

\begin{figure*}
\begin{center}
\ifpdf
    \subfigure[The example of a page with histograms.]{\includegraphics[ angle=90, width=0.65\paperwidth, keepaspectratio]{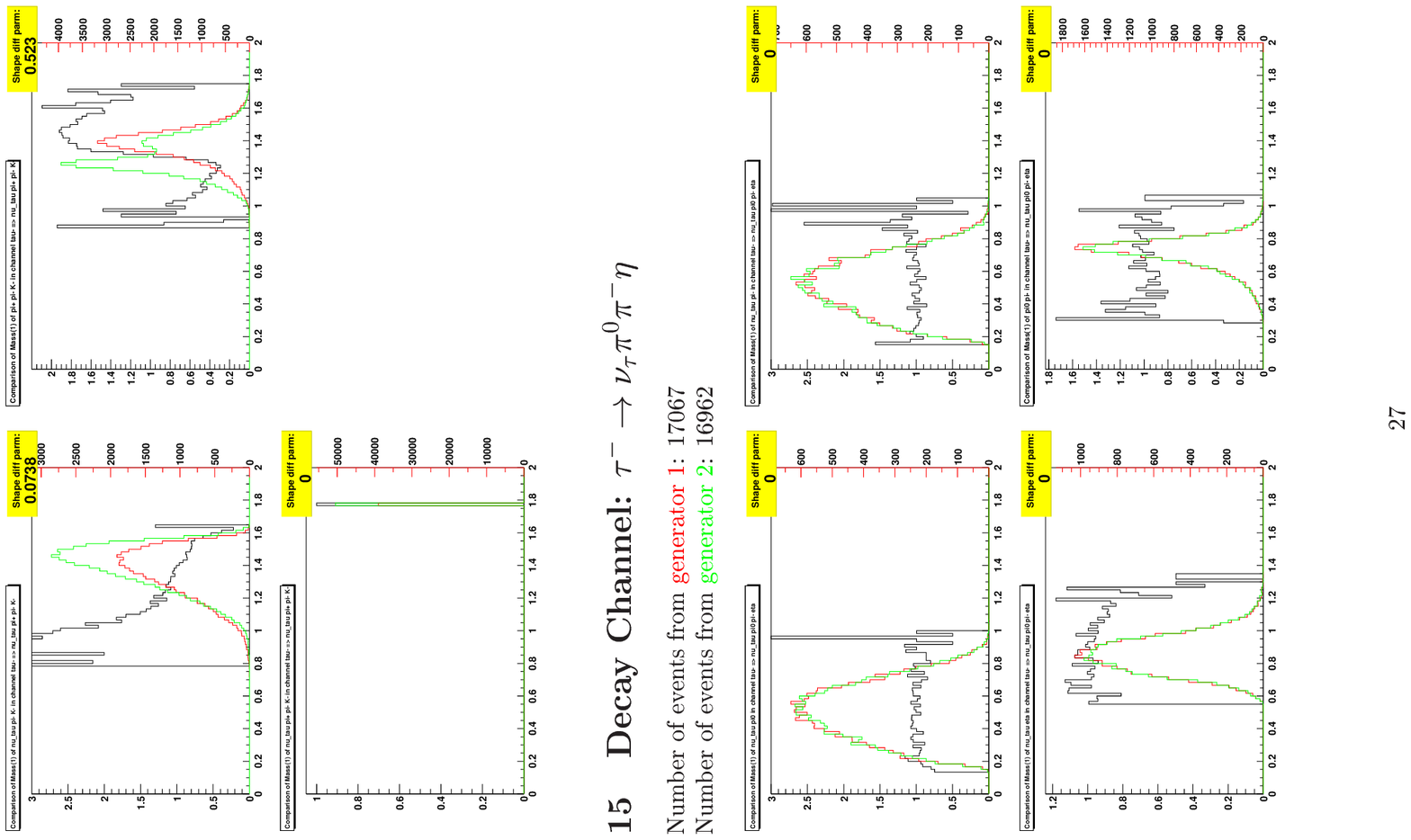}\label{subfig:singlepage}}\\
    \subfigure[The example of a histogram.]{\includegraphics[ width=0.45\paperwidth,keepaspectratio]{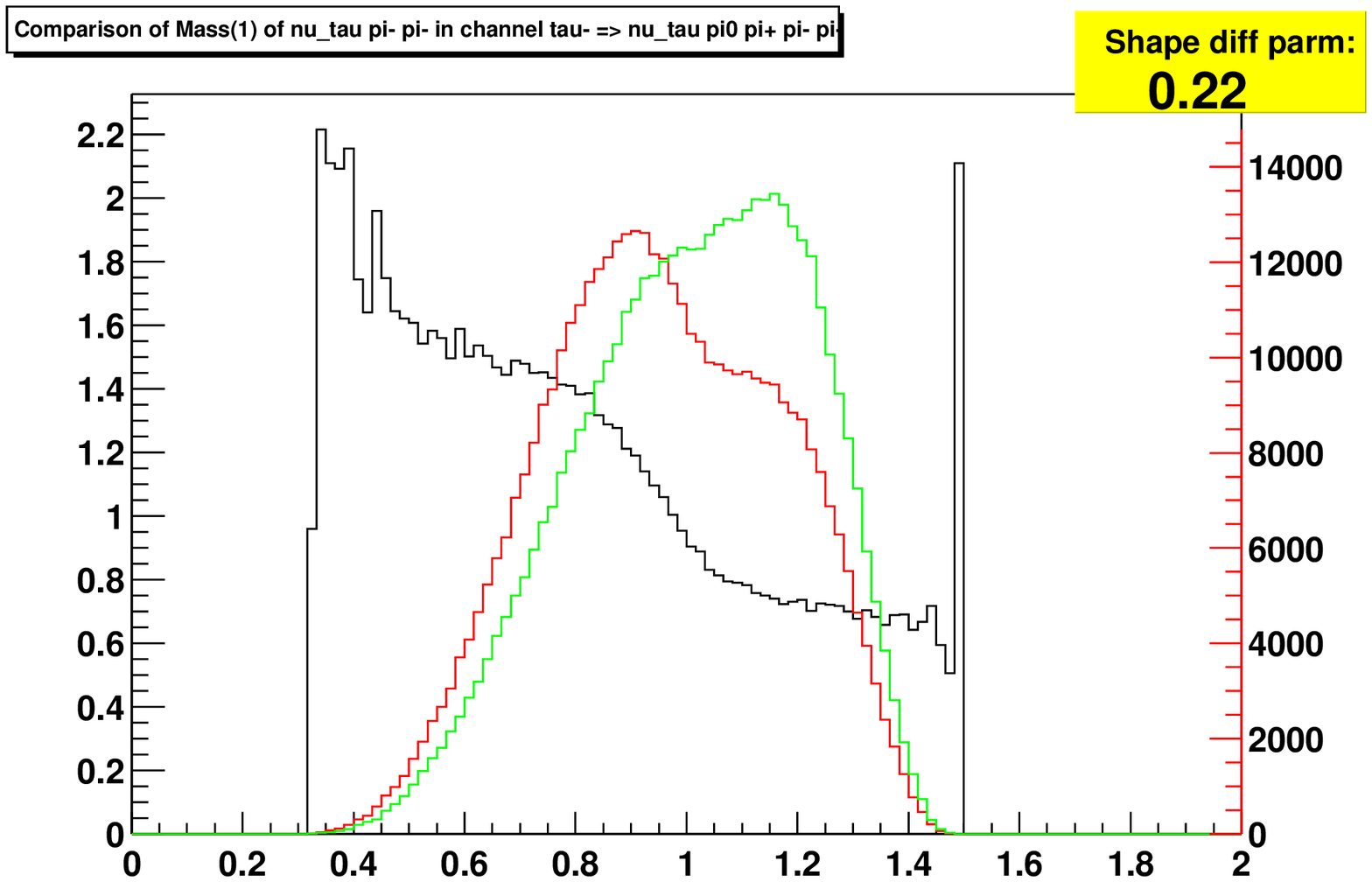}\label{subfig:singleplot}}
\else
    \subfigure[The example of a page with histograms.]{\includegraphics[  bb=100 150bp 500bp 750bp,
      clip,angle=90, width=0.65\paperwidth, keepaspectratio]{tester3}\label{subfig:singlepage}}\\
    \subfigure[The example of a histogram.]{\includegraphics[  bb=100 0 567bp 378bp,
      width=0.45\paperwidth,keepaspectratio]{tester4}\label{subfig:singleplot}}
\fi
\caption{ Examples of plots produced at analysis step.}
\label{fig-Tester}
\end{center}
\end{figure*}

%%%%%%%%%%%%%%%%%%%%%%%%%%%%%%%%%%%%%%%%%%%%%%%%%%%
Data files  {\tt mc-tester.root}, referred in
%explained in
 the previous section are used to produce 
a booklet  (Fig.\ref{fig-booklet1} and Fig.\ref{fig-Tester}) - a final results of the  {\tt MC-TESTER} action.
For this purpose the directory {\tt analyze/} is prepared. Unlike the rest of 
{\tt MC-TESTER}, the analysis code
is not stored in the {\tt MC-TESTER}'s libraries, but in  a set of {\tt C++/ROOT} 
macro routines.
These routines make use of the {\tt MC-TESTER} libraries and expect 
them in  directory {\tt lib/} .

The two data files from the generation step are expected to be found in the {\tt analyze/prod1/}
and  {\tt analyze/prod2/} directories respectively (these locations and many more
aspects of the analysis may be changed using the {\tt SETUP.C} file, see section \ref{SETUP.C}). 

A simple {\tt make} command in the {\tt analyze/} directory needs to be executed.
As a result,  the Postscript file {\tt tester.ps} with  the  complete booklet is produced.

The booklet consist of:
\begin {itemize}
\item The title page giving details of performed tests, ID of a tested particle,
      names and a short description of the two programs under test, numbers of generated events and 
      numbers of decay channels found in the two runs; see Fig. \ref{subfig:firstpage}.
\item The table of decay channels found with branching ratios from the two runs and  maximum 
      of the {\tt Shape Difference Parameter} for all  histograms 
defined for the channel; 
      see Fig. \ref{subfig:channelstable},
\item
      The table of content indicating a page number where plots for a given channel start.
\item Plots of histograms of all  invariant masses for all 
      matching decay channels found in the two runs; see  Fig. \ref{subfig:singlepage}.
\end{itemize}

Along with the booklet, a {\tt ROOT} output file {\tt mc-results.root}
is generated. It may be analyzed further,
by the advanced user.
A complete set of all plots of the booklet
(in the {\em .eps} format) is also stored in  
 subdirectory {\tt booklet/}~.

\subsection{Description of a single plot}

The plots such as in Fig. \ref{subfig:singleplot} are  the main part 
of the booklet. They
allow for the visual estimation  how 
a certain mass distribution compares between the two programs.
There are three histograms in each plot. Two histograms with the distributions 
from the first and the second generator, are plotted in red and green 
(or darker/lighter grey), and a histogram of the ratio
of the two (in black). 
 The red and green colors are used consistently throughout the
booklet to refer to the generators.

The left axis on the plot refers to the histogram of the ratio. This histogram is obtained
from  the division of the normalized histograms, 
therefore if the shapes of the two mass histograms match (even though they differ in
statistic), the division histogram will be up to statistical fluctuations
matching  flat distribution at  1.0~.
The division histogram is the first, visual comparison test.

The right axis of the plot refers to the contents of the mass histograms, therefore represents
the number of entries in every  single mass bin.
These histograms are not normalized, and  large differences in statistical samples will immediately 
show up. 
One will observe
that the histogram with smaller number of entries will occupy  small (negligible) part  of the plot. 
The range on the right axis is chosen such, that 
both histograms will show up in full.

The total number of events found for a given channel in the runs 
of the two generators is printed in the booklet, just at the beginning of 
a section including plots for a  channel (see Fig. \ref{subfig:singlepage}).

In the top-right corner of the plot (on yellow background), 
{\tt SDP} (the {\tt Shape Difference Parameter})
is printed.

%%%%%%%%%%%%%%%%%%%%%%%%%%%%%%%%
\section{{\tt Shape Difference Parameter} calculation algorithms}
\label{sec:SDP-algorithms}
%%%%%%%%%%%%%%%%%%%%%%%%%%%%%%%%
An analysis performed by {\tt MC-TESTER} consists of comparisons of distributions 
in every possible invariant mass constructed over every 
(sub-)set of decay products of every decay channel. 
Already in our example of the comparison of the decays of the $\tau$ lepton,
we  encounter  30 distinct decay modes. 
Realizing that for a typical (5 body) decay channel the algorithm defines 
26 distinct distributions, we may be faced up with an analysis of thousands of plots.
%%If we realise that  for typical
%% (5 body) decay  channel, the algorithm defines 26 distinct distributions, we must accept that the complete 
%comparison of the two programs lead to a thousand or more plots.
It is therefore convenient
to define a single measure (which we call the {\tt Shape Difference Parameter} {\tt (SDP)}),  measuring 
the difference between the same distributions from the two programs.
Later,
for every  analyzed channel, 
the maximum of {\tt SDP} over all  pairs of distributions, from the two programs
can be found
and printed in the review table of our package (in the last column in the Fig.\ref{subfig:channelstable}). 

We admit, that we were unable to find any universal, reasonable choice for {\tt SDP}.
Depending on  the user's particular need, the choice of an  algorithm
will be different. Also, shapes of the distributions (which depends on the 
decaying channel)
may affect the choice. 
On the other hand, the main purpose of {\tt SDP} is to draw user attention to decay channel 
where the differences of predictions from the two programs differ maximally. 
That is why, we believe, the choice 
of  {\tt SDP} in many cases will not be of prime importance.
At present, in our package we provide  three options, which we have found useful in  work on
the {\tt TAUOLA} Monte Carlo. For  unsatisfied user, they will
serve, at least, as  examples to develop a new one.
Let us list available  tests:
\begin{itemize}
\item  {\tt MCTest01} $-$ exclusive surface (subsection \ref{sec:MCTest01}), 
\item  {\tt MCTest02} $-$ non-uniformity  of the   histograms ratio (subsection \ref{sec:MCTest02}),
\item  {\tt MCTest03} $-$ Kolmogorov compatibility test (subsection \ref{sec:MCTest03}).
\end{itemize}
In the following subsections,   we provide intuitive definitions, 
followed by the detailed ones. Some illustrative examples, helpful to understand differences  
can be found in subsections \ref{sec:Results1} and \ref{sec:Results2}.

\subsection{How to set a particular test}
%\hspace{6.3mm}
The source code for the three tests included with {\tt MC-TESTER} are placed in {\tt src/} directory,
and compiled in the {\tt libMCTester} library.

The choice of the appropriate code is done in the file {\tt SETUP.C} of the directory {\tt analyze/}~.

To employ one of the provided tests, it is sufficient to specify it (by its name) in the {\tt SETUP.C}
file, i.e.

{\tt Setup::user\_analysis = MCTest02; }

The user may provide a custom analysis code in form of a {\tt ROOT} macro 
(refer to section \ref{sec:user-defined-sdp}).
We recommend putting the code in a single file in the directory {\tt analyze/} . An example of the 
user-provided code may be found in {\tt analyze/MyAnalysis.C}. 
%However, if the user provides a custom code, it needs to be loaded and interpreted by {\tt ROOT} at first.
One performs this step using the following code in the {\tt SETUP.C} file:

{\tt gInterpreter->LoadMacro("./MyAnalysis.C"); }

The full path name to the analysis code may be specified as a parameter (in this case: 
the file {\tt MyAnalysis.C} from the current directory.

One should note that loading the macro file does not automatically select the test function
to be used by {\tt MC-TESTER}. After loading the macro, one needs to specify explicitly the
name of the function to be used, as directed above. Therefore, one may have a set of various
test functions in a single file and then select them by name.

As an example, assume we have the {\tt MyTests.C} file with the following functions defined in it:
\begin{itemize}
\item {\tt double CompatibilityTest( TH1D{*},TH1D{*}); }
\item {\tt double MyMCTest( TH1D{*},TH1D{*}); }
\item {\tt double AreaTest( TH1D{*},TH1D{*}); }
\end{itemize}
To load this macro file and select the {\tt MyMCTest} function, one needs to put the following lines
in the {\tt SETUP.C} file:

{\tt gInterpreter->LoadMacro("./MyTests.C"); }

{\tt Setup::user\_analysis = MyMCTest; }

\subsection{{\tt MCTest01} $-$  exclusive surface }
\label{sec:MCTest01}

This test determines the size of the area under
the two compared, normalized to unity histograms, which is not simultaneously
under the two of them. Estimates of the statistical fluctuations are subtracted
bin-by-bin.
The test gives 
{\tt SDP} equal to $ 0$  
when the compared histograms are statistically 
identical, 
and {\tt SDP} equals $ 2$ when the histograms are completely disjoint.

\subsubsection{Detailed definition}
For every bin of the two histograms (of the equal number of bins) 
we calculate, 
 
\begin{equation}
{ \Delta _i = |_1N_i~ -~ _2N_i|~ } ;\; _1N_i~={_1n_i~ \over {\sum_j} \; {_1n_j}};\; _2N_i~={_2n_i~ \over {\sum_j} \; {_2n_j}};
\label{modul}
\end{equation} 
where $_1n_i$ and $_2n_i$ denote  the content of the bin $i$ respectively of the histogram no.1 and no.2, the 
 $\Delta _i$ -- the bin content difference is prone to statistical fluctuations.
The standard deviation of the bin contents equals 
%\begin{equation} 
% _{1,2}\sigma_i={_{1,2}N_i \over \sqrt{{_{1,2}n_i}(1- {_{1,2}N_i})   }}.
%\label{moduls}
%\end{equation} 
\begin{equation} 
 _{1,2}\sigma_i={_{1,2}N_i \cdot \sqrt{\frac{(1- {_{1,2}N_i})}{{_{1,2}n_i}} }}.
\label{moduls}
\end{equation} 
As in the case of the tests of the Monte Carlo programs, generation of sufficiently large 
sample is usually 
not a problem, to avoid statistical fluctuations, we will take 
instead of  $\Delta _i$,
\begin{equation}
{\tilde{\Delta }_i = \left\{ \begin{array}{r@{\quad: \quad}l}
     \Delta _i - \kappa \;{_1\sigma _i} - \kappa \;{_2\sigma _i}& 
     \Delta _i - \kappa \;{_1\sigma _i} - \kappa \;{_2\sigma _i}    \geq 0 \\ 
0 &  \Delta _i - \kappa \;{_1\sigma _i} - \kappa \;{_2\sigma _i}     <   0. 
 \end{array} \right. }
\label{max}
\end{equation} 

 Finally:
\begin{equation} 
{\tt SDP} = \sum_i \; \tilde{\Delta }_i,
\label{SDPa}
\end{equation} 
and as default we take  $\kappa=3$.
%*******************************************************************
\begin{figure}[!ht]
\setlength{\unitlength}{0.1mm}
\begin{picture}(1600,800)
\put( 375,750){\makebox(0,0)[b]{\large }}
\put(1225,750){\makebox(0,0)[b]{\large }}
\put(300,100){\makebox(0,0)[lb]{\epsfig{file=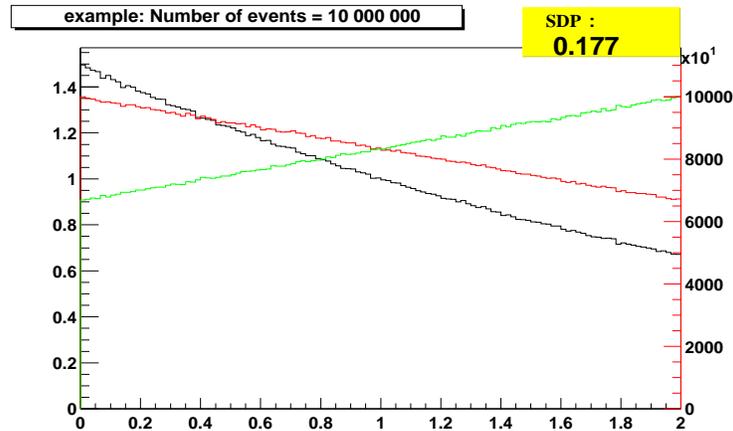,width=100mm,height=60mm}}}

\end{picture}
\caption
{\it Comparison of two histograms with 
{\tt SDP} $=0.177~$ (in the limit of infinite samples 
{\tt SDP}  would equal $0.20$).  }
\label{rys1}
\end{figure}
%************************************************************************* 
%*******************************************************************
\begin{figure}[!ht]
\setlength{\unitlength}{0.1mm}
\begin{picture}(1600,800)
\put( 375,750){\makebox(0,0)[b]{\large }}
\put(1225,750){\makebox(0,0)[b]{\large }}
\put(300,100){\makebox(0,0)[lb]{\epsfig{file=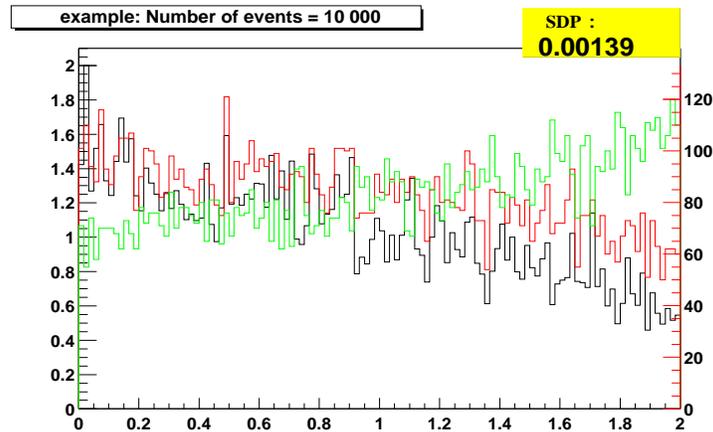,width=100mm,height=60mm}}}

\end{picture}
\caption
{\it 
Comparison as in the previous figure, but samples $1000$
times smaller.
{\tt SDP}~$=0.00139~$ now.}
\label{rys2}
\end{figure}
%*************************************************************************
 %*******************************************************************
\begin{figure}[!ht]
\setlength{\unitlength}{0.1mm}
\begin{picture}(1600,800)
\put( 375,750){\makebox(0,0)[b]{\large }}
\put(1225,750){\makebox(0,0)[b]{\large }}
\put(300,100){\makebox(0,0)[lb]{\epsfig{file=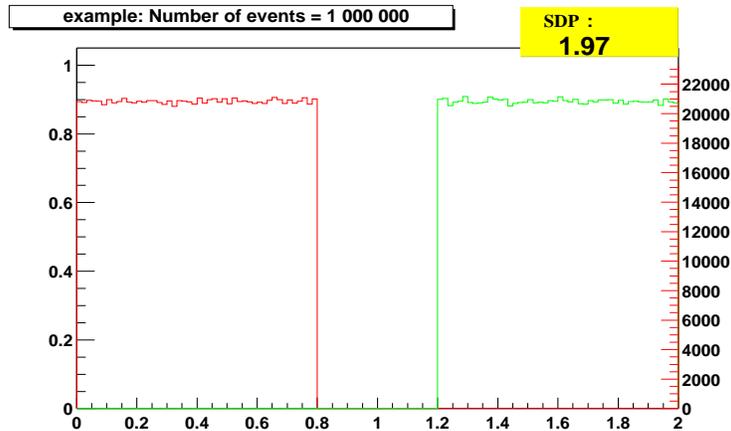,width=100mm,height=60mm}}}

\end{picture}
\caption
{\it Two completely disjoint histograms. 
{\tt SDP} $=1.97$.  }
\label{rys3}
\end{figure}
%*************************************************************************
 %*******************************************************************
\begin{figure}[!ht]
\setlength{\unitlength}{0.1mm}
\begin{picture}(1600,800)
\put( 375,750){\makebox(0,0)[b]{\large }}
\put(1225,750){\makebox(0,0)[b]{\large }}
\put(300,100){\makebox(0,0)[lb]{\epsfig{file=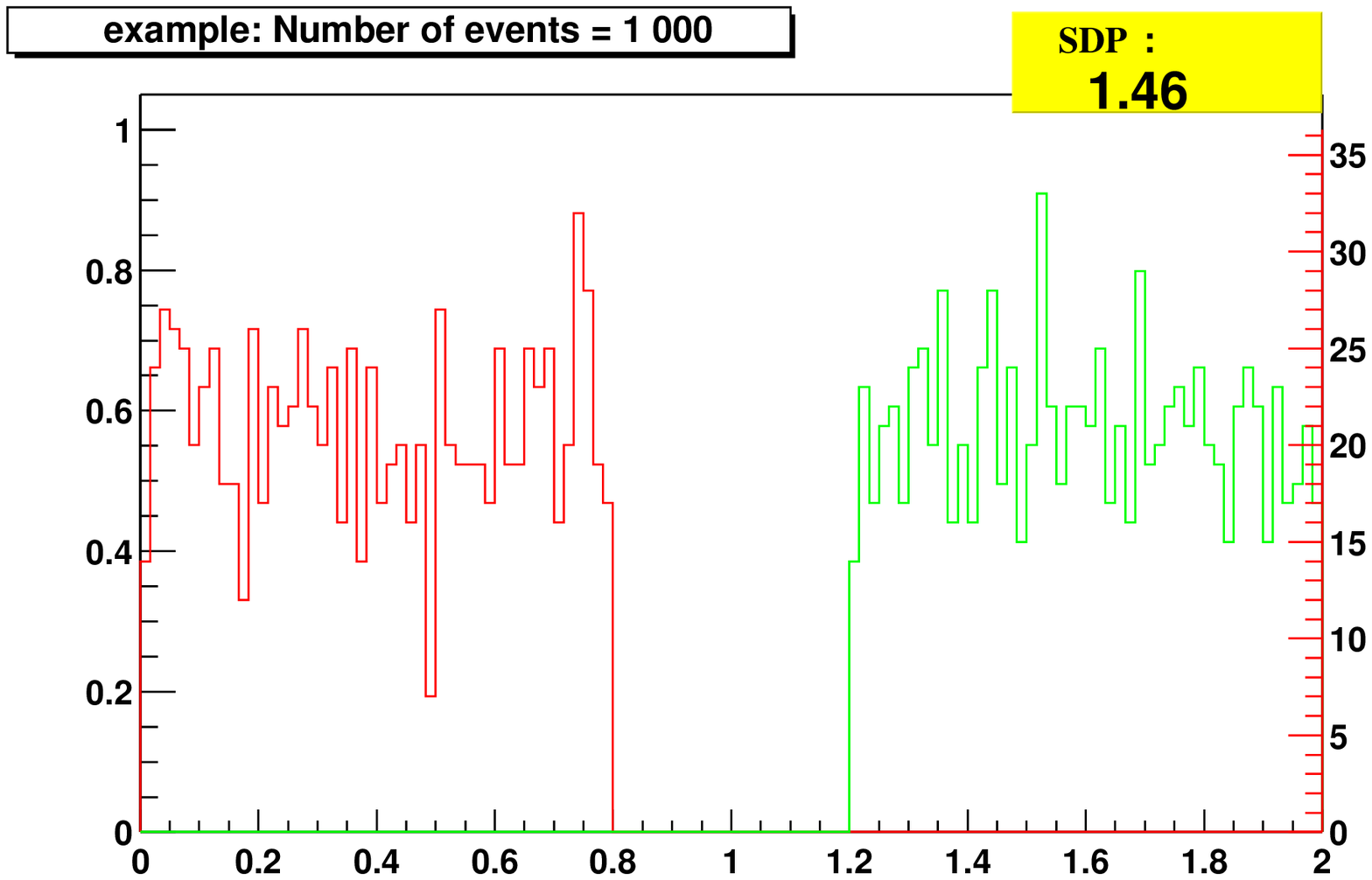,width=100mm,height=60mm}}}

\end{picture}
\caption
{\it Two completely disjoint histograms.
{\tt SDP} $=1.46$, samples 
  $1000$ times smaller than in the previous figure.}
\label{rys4}
\end{figure}
%*************************************************************************

\subsubsection{Results and interpretation of {\tt MCTest01} }
\label{sec:Results1}
The geometrical principle
of the test is to estimate an exclusive part of the area under the 
two (normalized to $1$) histograms. That is, this  part under each of them, which
is {\it not} simultaneously 
under the another. A convenient solution 
to put a result and a statistical error into a single number requires some 
modification.
We choose simply to subtract, bin by bin, the statistical error for the bin
contribution to ${\tt SDP}$ as given in formula (\ref{max}).  
 With the increasing  $\kappa $  the test 
will be less sensitive to the statistical fluctuations  of the compared histograms,
and at the same time it will tend to more and more 
underestimate ${\tt SDP}$. 
In the limit of infinite samples the results will 
be independent of $\kappa $.

In  Fig. \ref{rys1} we compare two  histograms of $10^7$ entries, generated from the 
analytic distributions  
of the disjoint  surface of  $0.2$ .  For this we obtain 
{\tt SDP}=0.177. {\tt SDP} is slightly underestimated.
In  Fig. \ref{rys2} for the histograms generated from the same analytic distributions, 
but with   $10^4$ entries we obtain  {\tt SDP}=0.00139. This is 
not a surprise because of the small statistics. With our assumption
of
$\kappa= 3 $ the histograms are found to be almost statistically equal. 
If our choice were $\kappa=1$,  the 
test would return {\tt SDP}=0.0746. 
The large numerical size
of  {\tt SDP} is originated from  statistical fluctuation.  

In our second example,  we take  two 
analytic distribution which are completely disjoint and we generate from them
the histograms (Fig. \ref{rys3}) of the samples of $10^{6}$ events.
{\tt SDP}=1.97 in this case.
Even for such a high  statistic and the clear separation, our test does not provide
{\tt SDP}=2. 
If, as in  Fig. \ref{rys4}, we take  $1000$ events, {\tt SDP}=1.46.

Note that our test, in general, provides the larger and larger 
{\tt SDP} with the increasing samples; on the other hand, {\tt SDP}
would decrease  with the increasing number of bins and  the constant statistics. 

\subsection{{\tt MCTest02} $-$ non-uniformity  of the   histograms ratio}
\label{sec:MCTest02}

The aim of this test is to measure, how far from a constant is a
ratio  of two histograms.   
The difference  of this test with respect to the previous one
is, that it weights equally all bins, and is not focused on the most 
populated bins as in the previous case.
The test  returns {\tt SDP}=0 
when the compared histograms are statistically 
identical. 
In general case, in the limit of infinitely large samples, it returns the
surface between two lines: the ratio of the two (normalized) histograms 
(see e.g. the black line and the left scale on Fig.\ref{fig-Tester}(d)) and
the constant line at 1.

\subsubsection{Detailed definition}

The test goes bin-by-bin. For every bin  we calculate first $_{1,2}N_i$
and $_{1,2}\sigma_i$ as given in formulas (\ref{modul}) and (\ref{moduls}) .
Later, to remove statistical fluctuations, we shift the results toward each other
\begin{equation}
{\tilde{_1N_i} = \left\{ \begin{array}{r@{\quad: \quad}l}
      _1N_i - \kappa \;{_1\sigma _i} & 
      _1N_i - {_2N}_i   \geq 0 \\ 
      _1N_i + \kappa \;{_1\sigma _i} &
      _1N_i - {_2N}_i    <   0, 
 \end{array} \right. }
\label{maxa}
\end{equation} 
and 
\begin{equation}
{\tilde{_2N_i} = \left\{ \begin{array}{r@{\quad: \quad}l}
      _2N_i - \kappa \;{_2\sigma _i} & 
      _2N_i - {_1N_i}   \geq 0 \\ 
      _2N_i + \kappa \;{_2\sigma _i} &
      _2N_i - {_1N_i}    <   0 .
 \end{array} \right. }
\label{maxb}
\end{equation} 

Finally, if the shift was larger than the original difference, we set
the contribution of the bin to zero, with the $\Theta_i$ function:
\begin{equation}
{\Theta_i = \left\{ \begin{array}{r@{\quad: \quad}l}
      0  &
      (_2N_i - {_1N}_i) (\tilde{_2N_i} - \tilde{_1N_i})   <   0 \\
      1  & 
      (_2N_i - {_1N}_i)(\tilde{_2N_i} - \tilde{_1N_i})   \geq 0 . 
 \end{array} \right. }
\label{maxc}
\end{equation} 

It is now straightforward to calculate

\begin{equation} 
{\tt SDP} = {1 \over N_{bin}}\sum_i \; 
\Bigl( {\tilde{_1N_i} \over\tilde{_2N_i}  }+ {\tilde{_2N_i} \over\tilde{_1N_i}  } -2\Bigr)\Theta_i,
\label{SDPb}
\end{equation} 

where $ N_{bin}$ denotes the number of bins in the histogram.

%*******************************************************************
\begin{figure}[!ht]
\setlength{\unitlength}{0.1mm}
\begin{picture}(1600,800)
\put( 375,750){\makebox(0,0)[b]{\large }}
\put(1225,750){\makebox(0,0)[b]{\large }}
\put(300,100){\makebox(0,0)[lb]{\epsfig{file=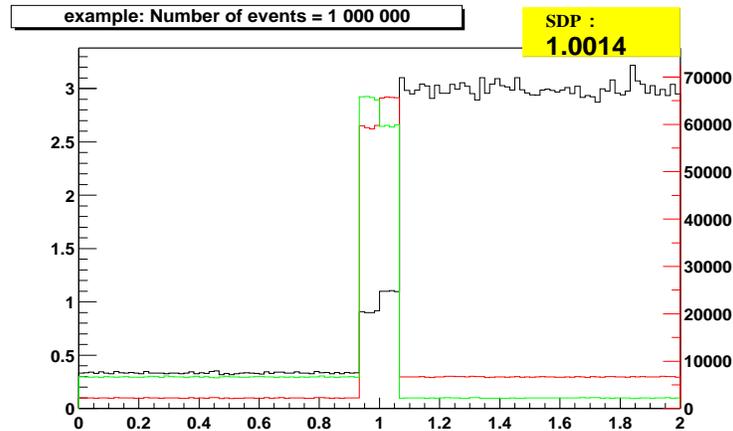,width=100mm,height=60mm}}}

\end{picture}
\caption
{\it For these two histograms (red and green) {\tt MCTest01} gives  
{\tt SDP}  $=0.4881$, and {\tt MCTest02} gives
{\tt SDP}  $=1.0014$.}
\label{rys.1i}
\end{figure}
%*************************************************************************
 %*******************************************************************
\begin{figure}[!ht]
\setlength{\unitlength}{0.1mm}
\begin{picture}(1600,800)
\put( 375,750){\makebox(0,0)[b]{\large }}
\put(1225,750){\makebox(0,0)[b]{\large }}
\put(300,100){\makebox(0,0)[lb]{\epsfig{file=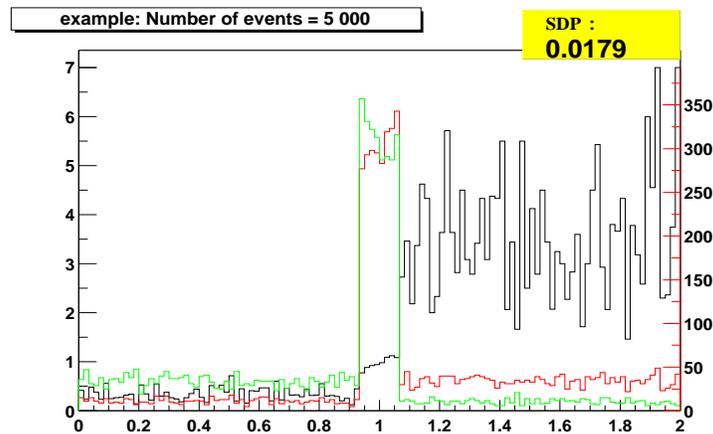,width=100mm,height=60mm}}}

\end{picture}
\caption
{{\it For these two histograms  {\tt MCTest01} gives  
{\tt SDP}  $=0.0235$, and {\tt MCTest02} gives
{\tt SDP}  $=0.0179$. Statistic is 200 times
smaller than in} Fig.\ref{rys.1i}.}
\label{rys.2i}
\end{figure}
%*************************************************************************
\subsubsection{Results and interpretation of {\tt MCTest02}}
\label{sec:Results2}
To better visualize the difference between {\tt MCTest01} and {\tt MCTest02}
 we provide Figs. \ref{rys.1i} and \ref{rys.2i}. 
The distributions in  
Fig. \ref{rys.1i} give smaller   {\tt SDP} from {\tt MCTest01} than from 
{\tt MCTest02}.
 Fig. \ref{rys.2i} shows the  histograms  generated 
from the same analytical distribution, but with the $200$ times lower
statistics. 
The situation is now reversed and {\tt SDP}  is larger 
with {\tt MCTest01}. This is because in the less populated bins, even though
the two distributions differ by a sizable factor, the statistics is  small and
the differences drop out as  statistically not significant.
\subsection{{\tt MCTest03}$-$ Kolmogorov compatibility test}
\label{sec:MCTest03}
In some applications, such as checks of proper installation into broader environment
of a code for the particle decay,
we may be interested if indeed after installation, all distributions
remained unchanged. For that purpose  Kolmogorov
compatibility test is suitable. It calculates a probability $p$ that the two histograms
have the same shape. As this test is  implemented and documented 
as a standard function of {\tt ROOT} \cite{root1,root2}, we will use it for the time being, 
and skip a description as well. 
We simply take {\tt SDP}=$1-p$. 
Test returns {\tt SDP}=$0$ if two distributions are identical, 
in other cases it returns the probability of being different.  
%    Although Kolmogorov  test runs correctly, we recall its invocation as detailed
%example.

%%%%%%%%%%%%%%%%%%%%%%%%%%%%%%%%%%%%%%%%%%%%%%%%%%%%%%%%%%%%%%%%%%%%%%%%%%%%%%
%%%%%%%%%%%%%%%%%%%%%%%%%%%%%%%%%%%%%%%%%%%%%%%%%%%%%%%%%%%%%%%%%%%%%%%%%%%%%%
\section{Package organization}
\label{sec:package-organization}
%%%%%%%%%%%%%%%%%%%%%%%%%%%%%%%%%%%%%%%%%%%%%%%%%%%%%%%%%%%%%%%%%%%%%%%%%%%%%%
%%%%%%%%%%%%%%%%%%%%%%%%%%%%%%%%%%%%%%%%%%%%%%%%%%%%%%%%%%%%%%%%%%%%%%%%%%%%%%
%In the following let us explain in detail organization of the package,
%how different parts of the code are distributed over the tree of subdirectories
%and what is their function for different modes of package operation such as its
%installation, phase of event generation and analysis.

This section contains technical details concerning {\tt MC-TESTER} and should be 
used as a quick reference.
Further details may be found in the Appendix and in files placed 
in the {\tt doc/} subdirectory.
%%%%%%%%%%%%%%%%%%%%%%%%%%%%%%%%%%%%%%%%%%%%%%%%%%%%%%%%%%%%%%%%%%%%%%%%%%%%%%
\subsection{Directory tree}
\label{dir-struct}
%%%%%%%%%%%%%%%%%%%%%%%%%%%%%%%%%%%%%%%%%%%%%%%%%%%%%%%%%%%%%%%%%%%%%%%%%%%%%%

\begin{description}
\item [doc/] - contains documentation.
\item [examples-F77/] - includes example programs in {\tt F77}:
        \begin{description}
                \item [tauola/] - using the {\tt TAUOLA} generator;
                \item [pythia/] - using the {\tt PYTHIA/JetSet} generator;
                \item [pre-generated/] - results of generation with high statistics.
        \end{description}
\item [examples-C++/] - examples in {\tt C++} (none at the moment).
\item [analyze/] - analysis step is performed in this directory,
        the analysis code is contained in a set of \texttt{ROOT} macros:
        \begin{description}
                \item [prod1/] 
                \item [prod2/] -  contains data files ({\tt mc-tester.root}) with 
                                the results of the generation phase
                    produced by  the two compared generators should be put here;
                \item [booklet/] - is created during the analysis step. It contains the result
                histograms in the form of {\it .eps} files.
        \end{description}
\item [HEPEvent/] - includes universal {\tt C++} interface to {\tt F77} event records
        ({\it i.e.} {\tt HEPEVT},{\tt LUJETS},{\tt PYJETS}).
\item  [include/]  - {\tt links} to {\tt C++} include files.
\item  [lib/] - contains compiled libraries needed by {\tt MC-TESTER}.
        Both the static and dynamic libraries are provided.
        
\item [src/]  - contains the source code for {\tt MC-TESTER}.
\item [platform/]    -  platform-dependent support files;
                        currently only for {\tt Linux}.
\end{description}

\subsection{Libraries}
From a point of view of a programmer,
% who wants to use the tool, 
{\tt MC-TESTER} is seen as a set of two libraries:
{\tt libMCTester} and {\tt libHEPEvent}. These libraries may be found in 
the {\tt lib/} directory.

The library {\tt libMCTester} contains all the code needed by the generation step; it is also required
at the analysis step, i.e. it contains routines for the {\tt Shape Difference 
Parameter} estimator.
%calculation.

The {\tt libHEPEvent} library contains a unified {\tt C++} interface for various 
{\tt F77 HEP} Monte Carlo
event record standards \cite{HEPEventlib}. Its first implementation was realized during
work on {\tt photos+} \cite{photosplus}, a {\tt C++} implementation of {\tt F77} 
algorithm for QED radiative
corrections \cite{Barberio:1990ms,Barberio:1994qi}. 
In the current version  of the {\tt MC-TESTER}, it 
provides a unified access to the {\tt HEPEVT}
({\tt HEPEVT} with $4000$  {\tt DOUBLE PRECISION} matrices is used\footnote{
To tune the size or precision of the {\tt HEPEVT} or {\tt LUJETS/PYJETS}
 used in {\tt MC-TESTER}, to  match  your code,
see {\tt README}  in the {\tt include/} directory.})
, {\tt LUJETS} and {\tt PYJETS} 
standards, enabling {\tt MC-TESTER} to be used with variety of Monte Carlo 
event generators,
based on those
%various 
 event record standards. 

We intend to extend this interface 
(and therefore  {\tt MC-TESTER})
to serve various future event standards, used by {\tt F77}, {\tt C++} or {\tt Java} -based event generators.

The source code of {\tt libMCTester} is placed in the {\tt src/} directory; {\tt libHEPEvent} is stored
in the {\tt HEPEvent/} directory.

%%%%%%%%%%%%%%%%%%%%%%%%%%%%%%%%%%%%%%%%%%%%%%%%%%%%%%%%%%%%%%%%%%%%%%%%%%%%%%
\subsection{Format and syntax of the {\tt SETUP.C} file}
\label {SETUP.C}
%%%%%%%%%%%%%%%%%%%%%%%%%%%%%%%%%%%%%%%%%%%%%%%%%%%%%%%%%%%%%%%%%%%%%%%%%%%%%%

The {\tt SETUP.C} file is a \texttt{C++ ROOT} macro file, which controls 
the {\tt MC-TESTER}'s settings.
It is read and executed during initialization of both phases 
of a {\tt MC-TESTER} run: the generation and the analysis. 
The file needs to be put in the same directory in which a run is executed, i.e.
in the directory in which 
executable
is being run for the generation phase, and in the {\tt analyze/} directory in the case of analysis. The setup file needs to have a correct 
{\tt C++}
syntax. Although any code (acceptable by {\tt ROOT}) may be put into the file,
the main purpose of the file is to set up the {\tt MC-TESTER} parameters. Majority of 
these
parameters are stored in a static class called {\bf Setup}, 
thus the most common
lines inside the {\tt SETUP.C} file have the  form:
~\\
\begin{minipage}[c]{1.0\textwidth}
\texttt{Setup::SomeSetting=somevalue;}
\end{minipage}
The {\tt SETUP.C} file needs to begin with \char`\"{}\texttt{\{}\char`\"{} and end with
\char`\"{}\texttt{\}}\char`\"{} characters.
No definition of a function name is needed.
Comment characters that may be used inside {\tt SETUP.C} are the ones
allowed by {\tt C++}: \char`\"{}\texttt{//}\char`\"{} (two slashes) marks out that a string 
following them  is a comment; \char`\"{}\texttt{/{*}}\char`\"{}
starts a comment which may span over any number of lines, and needs
to be terminated by the \char`\"{}\texttt{{*}/}\char`\"{} string. All text contained
between these two markers is treated as a comment. Comments in the form of \char`\"{}\texttt{//}\char`\"{}
may be nested inside \char`\"{}\texttt{/{*} {*}/}\char`\"{} comments.
Each line that is not a comment, i.e. that is a {\tt C++} statement, needs to be ended by the semicolon \char`\"{}\texttt{;}\char`\"{}
character.

A trivial {\tt SETUP.C} file, may look as follows:

\begin{minipage}[c]{1.0\textwidth}
\texttt{\{ }~\\
\texttt{/{*}{*}{*}{*}{*}{*}{*}{*}{*}{*}{*}{*}{*}{*}{*}{*}{*}{*}{*}{*}{*}{*}{*}{*}{*}{*}{*}{*}{*}{*}{*}{*}{*}{*}{*}{*}{*}{*}/
}~\\
\texttt{/{*} This is SETUP.C file for MC-TESTER {*}/}~\\
\texttt{/{*}{*}{*}{*}{*}{*}{*}{*}{*}{*}{*}{*}{*}{*}{*}{*}{*}{*}{*}{*}{*}{*}{*}{*}{*}{*}{*}{*}{*}{*}{*}{*}{*}{*}{*}{*}{*}{*}/}~\\
\texttt{// Some dummy variable}~\\
\texttt{ double x=1.2345;}\\
\texttt{\}}
\end{minipage}

One may put any call to a standard {\tt C} library function inside the {\tt SETUP.C}
 file, i.e. one may freely use the \texttt{printf()} call or the {\tt C++} 
\texttt{cout<\,{}<} streams to output strings of the text to the screen.
Alternatively, one can read the content of any file
using the \texttt{fopen}, \texttt{fscanf}, \texttt{fclose} calls, etc.
Usually, there is no need to put \texttt{\#include} statements, as majority of them is already
preloaded  by the {\tt ROOT} interpreter. 

The most important settings are presented in the Table \ref{table:settings}.

\begin{table}[h]
\begin{tabular}{|c|c|c|c|}
\hline
    Variable & {\tt C++} Type & Default & Description \\
\hline
\hline
Setup::decay\_particle     &  int   &  15 ($\tau^{-}$)     & {\tt PDG} code of particle,  \\
                           &        &                      & which decays we analyze\\
\hline
Setup::EVENT              & HEPEvent{*} & HEPEVT & event record format: \\
                          &             &        & (\texttt{HEPEVT, LUJETS, PYJETS})  \\
\hline
Setup::gen1\_desc\_1 & char{*} & &  three lines of text\\
Setup::gen1\_desc\_2 & char{*} & &  describing \\
Setup::gen1\_desc\_3 & char{*} & &  the first generator\\
\hline
Setup::gen2\_desc\_1 & char{*} & & three lines of text\\
Setup::gen2\_desc\_2 & char{*} & & describing\\
Setup::gen2\_desc\_3 & char{*} & & the second generator\\
\hline
Setup::user\_analysis& double(*)      & (none) & function for {\tt SDP} \\
                     &  (TH1D*,TH1D*) &         & \\
\hline
Setup::nbins & int{[}20{]}{[}20{]} & 120 & number of bins in histograms\\
        &       & for all values &      \\
\hline
Setup::bin\_min & double {[}20{]}{[}20{]} & 0.0 & the lowest bin in histograms\\
        &       & for all values &      \\
\hline
Setup::bin\_max & double {[}20{]}{[}20{]} & 2.0 & the highest bin in histograms\\
        &       & for all values &      \\
\hline

\end{tabular}
\caption { The most important settings of {\tt MC-TESTER}. }
\label{table:settings}

\end{table}

The complete list of parameters may be found in the Appendix.
It may also be found in the {\tt doc/README.SETUP} file.

\subsection{User-defined {\tt Shape Difference Parameter} algorithms}
\label{sec:user-defined-sdp}

    In order to employ the user-defined analysis routine, one has to write a {\tt ROOT}
macro file in {\tt C++}, with a function which accepts pointers to 
two histograms of {\tt TH1D} class, and returns {\tt double}. 
For example:

~\\
\begin{minipage}[c]{1.0\textwidth}
{\tt double MyTest(TH1D *h1, TH1D *h2)\\
$\{$\\
double value=1.0 - h1->KolmogorovTest(h2);\\
return  value;\\
$\}$
}\\
\end{minipage}
An analysis function gets {\em copies} of histograms from two generators as
the {\tt h1} and {\tt h2} parameters, so it may perform any
modifications on them (i.e. normalizations, divisions, etc.).
The function should be saved 
in the directory {\tt analyze/} in the file {\tt MyAnalysis.C}. Its invocation 
has to be uncommented in {\tt SETUP.C} in {\tt analyze/}, and invocation to the
standard one has to be commented out. 
One may also specify the full pathname to the analysis code. For details,
see the Appendix.

%%%%%%%%%%%%%%%%%%%%%%%%%%%%%%%%%%%%%%%%%%%%%%%%%%%%%%%%%%%%%%%%%%%%%%%%%%%%
\subsection{How to make {\tt MC-TESTER} run with other generators}
\label{other-generators}
%%%%%%%%%%%%%%%%%%%%%%%%%%%%%%%%%%%%%%%%%%%%%%%%%%%%%%%%%%%%%%%%%%%%%%%%%%%%

\subsubsection{The case of {\tt F77} }
\label{other-gen:F77}
The {\tt MC-TESTER} routines may be called directly from the {\tt F77} code, though 
all the routines are written in {\tt C++}. "{\tt F77} wrappers" are 
provided to access transparently the {\tt MC-TESTER} routines.

As a starting point, one should follow 
{\tt examples-F77/tauola/tautest.F} \\ and
{\tt examples-F77/pythia/pythiatest.f}

From within a {\tt FORTRAN} code, one has the access to {\tt MC-TESTER} using
the {\tt MCTEST} subroutine, which accepts a single, integer parameter:
\begin{itemize}
\item $-1$ : initialization; must be called at the beginning
\item $~~0$ : generation step; should be called every time when {\tt HEPEVT}
        is  filled with a new event.
\item $~~1$ : finalization; closes output files, {\it etc.}
\item $~20$ : makes a printout of the currently used event record.
%{\tt HEPEVT} (similar to {\tt LULIST}).
\end{itemize}

Please refer to comments in the Appendix \ref{appendix.F77} or in the {\tt /doc/README.SETUP.F77} for the detailed description of
other {\tt F77} utility functions, which control the {\tt MC-TESTER} setup.

In order to use the {\tt MC-TESTER} routines, one needs to link two {\tt MC-TESTER}
 libraries and a subset of the {\tt ROOT} libraries. 
The {\tt MC-TESTER}'s libraries may be found in the {\tt lib/} directory.
There exist both the static ({\tt .a}) and dynamic ({\tt .so}) 
versions of the libraries:
    \begin{itemize}
    \item {\tt libHEPEvent.so} $-$ contains the {\tt C++} interface to {\tt HEP}event record structures
    \item  {\tt libMCTester.so} $-$ contains the {\tt MC-TESTER code}.
    \end{itemize}

The set of the {\tt ROOT} libraries needed by {\tt MC-TESTER} may be obtained 
by executing   

\begin{itemize}
\item  {\tt \$ROOTSYS/bin/root-config --libs} .
\end{itemize}

One could also add the following line to a {\tt Makefile}:
\begin{itemize}
\item {\tt ROOTLIBS := \$(shell \$(ROOTSYS)/bin/root-config --libs)} ,
\end{itemize}
then specify the linking of {\tt \$(ROOTLIBS)}.
It may also be required to link the {\tt F77} library: one should append
{\tt  -lg2c} at a linking step, and use {\tt g++} as a linker rather
  than {\tt g77}.

Before execution of a program linked with {\tt MC-TESTER}, one should 
prepare {\tt SETUP.C} file, and put it in the same directory as
executable file. For details, please refer to section \ref{SETUP.C}.

The results of the generation step are put in the file 
{\tt mc-tester.root} in the directory of the executable.
This output file should be moved (copied) to the {\tt MC-TESTER}'s {\tt analyze/prod1/}
or {\tt analyze/prod2/} directory to proceed with the analysis step.

\subsubsection{The case of {\tt C++} }
Although we currently do not provide any example in {\tt C++}, 
the infrastructure for
connecting {\tt MC-TESTER} to a {\tt C++} generator is already in place.

Settings for the {\tt MC-TESTER} parameters are done using a singleton
{\tt Setup} class.
All the settings in the {\tt SETUP.C} macro file, as described in the section \ref{SETUP.C}
refer to this object.

Inside the tested Monte Carlo analysis, it is sufficient to issue calls to the
following three routines:
\begin{itemize}
    \item{ \tt MC\_Initialize()}: initializes {\tt MC-TESTER}. All changes
        to the {\tt Setup} should be commenced before a call to this function is invoked.
    \item{ \tt MC\_Analyze(int particle)}: performs the analysis of
        the event record specified in the {\tt Setup::EVENT} variable; 
        {\tt particle} parameter should be the {\tt PDG} code of the decaying particle
        one wants to analyze\footnote{Please note that unlike in the case of the FORTRAN {\tt MCTEST(0)}
        function, one needs to specify the PDG code -- the code from {\tt Setup} is not
        passed automatically; i.e. one should use {\tt MC\_Analyze(Setup::decay\_particle);} }.
    \item{ \tt MC\_Finalize()}: writes the results to the output file.
\end{itemize}

One may also make use of the utility function
{\tt Setup::SetHistogramDefaults(int nbins, double min\_bin, double max\_bin)}
to configure the size and the range of histograms (\ref{appendix.S.SetHistDefaults}).

As in the case of {\tt F77} generators, one needs to link the {\tt MC-TESTER}'s libraries
and a subset of {\tt ROOT} libraries.
The {\tt HEPEvent} library \cite{HEPEventlib} is intended to serve as a 
universal interface to future event record standards in {\tt C++}. 
All event record standards included in the current version of {\tt HEPEvent}
library, i.e. {\tt HEPEVT}, {\tt LUJETS} and {\tt PYJETS} may directly
be used in user's {\tt C++} code.

%%%%%%%%%%%%%%%%%%%%%%%%%%%%%%%%%%%%%%%%%%%%%%%%%%%%%%%%%%%%%%%%%%%%%%%%%%%%%%
%%%%%%%%%%%%%%%%%%%%%%%%%%%%%%%%%%%%%%%%%%%%%%%%%%%%%%%%%%%%%%%%%%%%%%%%%%%%%%

%\section{Summary and future possibilities}

\section{Outlook}
\label{sec:outlook}
We have found that   {\tt MC-TESTER} is useful for some tests of 
 libraries of particles decays, but its tests are not complete
from the physics point of view. It also has some technical limitations.
In the following let us list these points, which can, in some cases, be fixed
in the future versions of   {\tt MC-TESTER}.

\begin{enumerate}
\item
The program constructs distributions out of stable decay products of 
the particle under study. It ignores intermediate states in the cascade
formed  from the decaying particle.
\item
The program does not analyze distributions in Lorentz invariants built with the 
help of totally antisymmetric (Levi-Civita) tensor. It is thus blind to some 
effects of parity non-conservation. 
\item
Information on the spin state of the decaying particle is usually not 
available in common blocks such as {\tt HEPEVT}.
To keep  {\tt MC-TESTER}  modular, and to avoid 
a multitude of options, we omit
effects of decaying particle polarization. 
%This is done automatically with our choice of distributions. 
Our choice of distributions is blind to these effects anyway.
\item
The main advantage of    
{\tt MC-TESTER} is that it can be used with `any' production generator in 
an automated way, providing a tool for quick tests. However, the final state 
event record has to be stored in one of the following common blocks:
{\tt HEPEVT, LUJETS, PYJETS} \cite{PDG:1998,Pythia} of {\tt FORTRAN}.
Further possibilities, in particular data structures of {\tt C++}, are not included 
at present.
\item
If multiplicity of the particular decay channel is very high and/or
there is a lot of decay channels, the program may
find it difficult to allocate memory. 
An analysis of a decay channel with 8 or more decay products produces
thousands of histograms, which causes data files to be huge and the analysis
step to be long.
In the current version, {\em the user is not warned} about possible problems. 
We observe
that a machine with a sufficient amount of memory may cope with an analysis of
large-multiplicity decay channels; however, the analysis process may take a long time
(tens of minutes on a 2~GHz machine). There may also be problems with producing a booklet
with a few thousands of histograms: one may run out of disk space.
Therefore we advise users to restrain from the analysis of high multiplicity decay channels
in the current version of {\tt MC-TESTER}. In the {\tt PYTHIA} example, we switch off the
decay channels with higher decay multiplicities (greater than 7).
\item
For some settings/types of the linker, all the input event records (the common blocks 
{\tt HEPEVT, LUJETS, PYJETS}) may be loaded simultaneously, even if only one of them is used.
To avoid problems with memory allocation, their size should be checked and adjusted to
 match declarations in the user's program and/or other libraries loaded. 
\item 
The present version of  {\tt MC-TESTER} uses the {\tt ROOT} package for the purpose of histograming, input/output, etc. 
Another similar package could in principle replace {\tt ROOT}.
\end{enumerate} 

Thanks to the interactions with the first users,
we envisage the extension of the {\tt MC-TESTER}
 with the following functionalities already in the next release:
\begin{itemize}
\item Allow automatic histogram re-binning (see variable {\tt nbins} table
\ref{table:settings})
at the analysis step. Any 
common integer divider for the  dimensions of the two compared
histograms will be allowed.  
\item
   A list of the particles to be treated as stable as well as a list of the decay products  
   not be taken into account (e.g.  $\pi^{0} \to \gamma \gamma$ decays) 
   will be introduced\footnote{At present {\tt MC-TESTER} constructs distributions from the final (stable) decay products of the analyzed particle.}.
\item Introduction of a new parameter to limit the depth of  a decay tree to be analyzed
   (e.g. at most, only secondary decay products of the 
   analyzed particle will be taken  and then treated as stable).
\end{itemize}

\subsection{Recent updates and extensions}
\label{subsec:recent_updates}

From the "to be introduced" options listed above, we have managed to implement the suppression of decays
of particles with certain PDG code (e.g. $\pi_0$) into present release (v.1.1)
of the {\tt MC-TESTER} (please refer to the Appendix \ref{option:SuppressDecay}
 for details).
    
The  {\tt MC-TESTER} has recently been proposed as the tool for the
Linear Collider  Workshop \cite{LC-Workshop} for comparisons of
Monte Carlo predictions for $2 \to n$; (n=4,6,8) body generators.
Some changes were necessary.

First, the possibility to process the program input was introduced.
The scattering event of the form $p_1 p_2 \to q_1 ... q_n$ can
be replaced by the decay-like chain : $P \to p_2 q_1 ... q_n$, where 
$ P = p_1 + 2 \times p_2 $ .
The first incoming particle (of momentum $p_1$) is replaced by the
$P$, the "mother" of all other particles, including second beam ($p_2$).
One obtains the event record with decay tree-like structure, i.e.
standard event record for {\tt MC-TESTER} use.
This extension is easily customizable, the user may switch it on
or modify using the C++ macro files: for details refer to 
README.EVENT-ANALYSIS and README.LC files in the {\tt /doc} directory.
With the help of this macro, the user may introduce other pre-processing
of the event record as well. Note, that in every case the event record
will only be modified locally, inside {\tt MC-TESTER}.

The second extension prepared for the Linear Collider Workshop allows to
quantify the difference of predictions of two programs for 
$p_1 p_2 $ ( or $p$ ) $\to$ {\em anything} process. This is done with the
help of two real numbers, which can be later analyzed if comparison of more
than two programs is performed.

The first number $T_1$ represents the difference due to branching ratios from
program A and B. It is calculated using the following formula:
\begin{equation}
    T_1 = \sum_i{max\Bigl(|Br_i^A-Br_i^B|-\kappa \sqrt{\sigma^2(Br_i^A)+\sigma^2(Br_i^B)} ; 0\Bigr)} ,
\end{equation}
(statistical fluctuations subtracted, in a similar way as in formula \ref{max}).
The second one:
\begin{equation}
    T_2 = \sum_i{\frac{Br_i^A+Br_i^B}{2}}(SDP_i^{max})
\end{equation}
consists of the weighted sum of maximal shape difference parameter calculated 
for all channels. The weight is given by the average of the branching ratios 
(calculated for $i$ channel ).

The values of $T_1$, $T_2$ are printed under the table of decay channels
(page 2 of the booklet).
They are also stored/appended to {\tt MC-TESTER.DAT} file in the analysis
directory for further use. 
The format of data in this file ( Comma Separated Values (CSV), 
containing descriptions of generators and the $T_1$, $T_2$ values) 
allows for easy processing by any data analysis tool or programming language.

For the example of the Workshop-like use see {\tt examples-F77/pythia.Lin-Col}
directory.
\section*{Acknowledgments}
We thank Thorsten Ohl for suggestions concerning functional extension of the
{\tt MC-TESTER} package.
 We thank Swagato Banerjee, Stanis\l aw Jadach and Wies\l aw P\l aczek  
for valuable remarks.

This work is partially supported by the BMBF (WTZ) project number POL~01/103.
One of the authors (T. P.)  would like  to thank the
``Marie Curie Programme'' of the European Commission  for a fellowship.

\include{appendix}

%%%%%%%%%%%%%%%%%%%%%%%%%%%%%%%%%%%%%%%%%%%%%%%%%%%%%%%%%%%%%%%%%%%%%%%%%%%%
%%%%%%%%%%%%%%%%%%%%%%%%%%%%%%%%%%%%%%%%%%%%%%%%%%%%%%%%%%%%%%%%%%%%%%%%%%%%
%\bibliographystyle{utphys_spires}
%\bibliographystyle{plain}
%\bibliography{TAUOLA-F}
%\bibliography{tester}
%\bibliographystyle{utphys_spires}
%\bibliographystyle{plain}
%\bibliography{Tester}

%%%%%%%%%%%%%%%%%%%%%%%%%%%%%%%%%%%%%%%%%%%%%%%%%%%%%%%%%%%%%%%%%%%%%%%%%%%%
%%%%%%%%%%%%%%%%%%%%%%%%%%%%%%%%%%%%%%%%%%%%%%%%%%%%%%%%%%%%%%%%%%%%%%%%%%%%

\providecommand{\href}[2]{#2}

\end{document}

%% file: appendix.tex
\appendix
\section{Appendix: {\tt MC-TESTER} setup and input parameters}
\label{appendix.A}

The values of the parameters used by {\tt MC-TESTER} are controlled using 
the {\tt SETUP.C} file. Some parameters may also
be controlled using {\tt FORTRAN77} interface routines 
(Section \ref{SETUP_in_F77}).
This provides a runtime control over all parameters, yet allowing the user
not to have  {\tt SETUP.C} at all.
One should note that {\tt SETUP.C} has always precedence over the values set 
using {\tt F77} code: it is always looked for in the execution directory.

Any parameter, not set using either of the methods, will have a reasonable 
default value,
which is quoted in the parameter's description below.

\subsection{\label{SETUP.C_FORMAT_AND_USE}Format and use of the {\tt SETUP.C} file}

Please refer to section \ref{SETUP.C}

\subsection{Definition of parameters in the {\tt SETUP.C} file}
\label{appendix.Setup-parameters}
There are three sets of settings inside {\tt MC-TESTER} to be distinguished: 
the ones
specific to the generation phase, the ones specific to the analysis phase and the ones
that are used in both phases\footnote{Some parameters from the generation 
> phase (i.e. the description of generators) are stored inside
> an output data file. However, again for reasons of runtime control, their 
> values may be altered at the analysis time using the {\tt SETUP.C} file in 
> the analysis directory.}. We describe them quoting the scope of their use.

\subsubsection{Setup::decay\_particle }

Type:~ int 

Scope:~ generation 

Default: 15 (\( \tau ^{-} \)) 

DESCRIPTION: the PDG code of a particle, which decay channels we want to analyze.

Example of use: 

\texttt{Setup::decay\_particle = -521; //analyze B0 decays.}

\subsubsection{Setup::EVENT }

Type:~ HEPEvent{*} 

Scope:~ generation 

Default: (HEPEVT) 

DESCRIPTION: the {\tt F77} event format used by generator. It must be supported by
the {\tt HEPEvent}
library. The possible values are: {\tt HEPEVT} (format: 4000 entries, 
double precision), {\tt LUJETS} (as in PYTHIA 5.7), {\tt PYJETS} (as in PYTHIA 6)

To change the format of the event record standards used by {\tt MC-TESTER} 
(the size of arrays
and the precision) please refer to the {\tt include/README} file.

Example of use:

\texttt{Setup::EVENT=\&LUJETS;}

\subsubsection{Setup::stage }

Type:~ int 

Scope:~ generation, analysis 

Default: - 

DESCRIPTION: Indicates whether this is a generation or analysis stage, and which
generator is being used: 0 = the analysis stage; 1 = the generation phase for the generator
1; 2 = the generation phase for the generator 2. \\This setting, %, in particular
 is responsible
for deciding which description specified by the {\tt SETUP.C} settings will be saved at
the generation phase. It is automatically set to $0$ at the analysis stage and needs to
be set by the user's program to 1 or 2 at the generation phase (it is automatically reset to 1 if 0 occurs
at the generation phase).\footnote{One of the trick in which it may be introduced to two
versions of the code, may be observed in the {\tt TAUOLA} example program, where the \char`\"{}template\char`\"{}
{\tt F77} code is preprocessed to produce two versions of the source codes, each having
a different stage set by means of the {\tt F77} interface 
(see {\tt doc/README.SETUP.F77} for details).}

Please note that at the analysis step one may freely replace the data files from the generation steps.
The description in the {\tt SETUP.C} file referring to the generator 1 will be applied to the data file
in the subdirectory {\tt analyze/prod1/} and the ones that refer to the generator 2 will apply to 
the data in {\tt analyze/prod2/}.

Example of use: (none)

\subsubsection{Setup::gen1\_desc\_1 , Setup::gen1\_desc\_2, Setup::gen1\_desc\_3}

Type:~ char{*} 

Scope:~ generation, analysis 

Default: {[}some default text with warnings{]} 

DESCRIPTION: Up to three lines containing the description of the first of used generators.
These lines will appear on the first page of the booklet produced at the analysis
step. Any proper \LaTeX{} sequences may be introduced inside, however one needs
to note the fact, that \textbackslash{} (slash) sign is interpreted as an escape
character in {\tt C/C++}, so one needs to use \textbackslash{}\textbackslash{} (double
slash) to introduce \char`\"{}\textbackslash{}\char`\"{} into output. See the example
of the use below. When specified at the generation step, this text will be saved in
the output file. If the corresponding {\tt SETUP.C} file will not alter these 
variables at the analysis phase, the text will appear on the first page of 
the booklet. However if these variables are being set in {\tt SETUP.C} 
in the analysis phase, they will have
a precedence over the ones stored in the generation files, so one may control the
text appearing in the booklet without need to re-run the generation process.

Example of use:

\texttt{Setup::gen1\_desc\_1=\char`\"{}PYTHIA version 5.7, JetSet version 7.4;
p-p at 14~TeV, \$Z\textasciicircum{}0\$ production\char`\"{}; }

\texttt{Setup::gen1\_desc\_2=\char`\"{}\$Z\textasciicircum{}0\$ decays to \$\textbackslash{}\textbackslash{}tau\textasciicircum{}-\$
exclusively. No \$\textbackslash{}\textbackslash{}pi\$ decays, No ISR/FSR.\char`\"{};}

\texttt{Setup::gen1\_desc\_3=\char`\"{}\{\textbackslash{}\textbackslash{}tt
You may replace this text in SETUP.C file.\}\char`\"{};}

\subsubsection{Setup::gen2\_desc\_1, Setup::gen2\_desc\_2, Setup::gen2\_desc\_3}

Type:~ char{*} 

Scope:~ generation, analysis 

Default: {[}some default text with warning{]} 

DESCRIPTION: The same as above, for the second generator. 

Example of use:

\texttt{Setup::gen2\_desc\_1=\char`\"{}TAUOLA LIBRARY: VERSION AA.BB\char`\"{};}

\texttt{Setup::gen2\_desc\_2=\char`\"{}.............................\char`\"{}; }

\texttt{Setup::gen2\_desc\_3=\char`\"{}\{\textbackslash{}\textbackslash{}tt
You may replace this text in SETUP.C file in analysis dir.\}\char`\"{};}

\subsubsection{Setup::result1\_path}
\label{appendix.S.result1-path}

Type:~ char{*} 

Scope:~ generation 

Default: (set automatically to \char`\"{}./mc-tester.root\char`\"{}) 

DESCRIPTION: Sets the path and a file name of the data file produced at the generation
phase. Note that the path (absolute or relative) {\bf AND} the filename needs
to be specified. Also take into account that the analysis step requires generation
output files to be placed in certain directories 
({\tt analyze/prod1}, {\tt analyze/prod2})
and to be named \char`\"{}{\tt mc-tester.root}\char`\"{}. 

Example of use:

\texttt{Setup::result1\_path = \char`\"{}/a/path/to/results/mc-tester.root\char`\"{}}

\subsubsection{Setup::result2\_path}
\label{appendix.S.result2-path}

Type:~ char{*} 

Scope:~ generation 

Default: (set automatically to \char`\"{}./mc-tester.root\char`\"{}) 

DESCRIPTION: The same as above, for the second generator. 

Example of use:

\texttt{Setup::result2\_path = \char`\"{}../prod/mc-tester.root\char`\"{}}

\subsubsection{Setup::order\_matters}

Type:~ int 

Scope:~ generation 

Default: 0 

DESCRIPTION: This switch (values 0 or 1) specifies the behavior of a routine
which searches for decay channels inside event records. By default (value: 0),
%the same particles are not recognisable, and 
the order in which decay products
are written in an event record is not important. 
However for debugging purposes it may be
useful to distinguish the order used by two generators. In that case, for example,
{[}pi- pi0 pi+{]} will be other decay channel than {[}pi+ pi- pi0{]}. At default
behavior, when the order is not taken into account, particles are sorted according
to their PDG code, and regrouped in such a way that antiparticles stay just
after corresponding particles. Example of use: 

\texttt{Setup::order\_matters = 1; }

\subsubsection{Setup::nbins}

Type:~ 2-D array: int{[}MAX\_DECAY\_MULTIPLICITY{]}{[}MAX\_DECAY\_MULTIPLICITY{]} 

Scope:~ generation 

Default: 120 

DESCRIPTION: Setup::nbins{[}n{]}{[}m{]} specifies the number of bins in
histogram of m-body invariant in the n-body decay mode. 
Look at the example below to get it clarified.
For setting default values to the whole range, use the {\tt Setup::SetHistogramDefaults()}
function described below.\\ The maximum number of decay products is equal 
to {\tt MAX\_DECAY\_MULTIPLICITY$-$1},
because arrays in {\tt C/C++} are indexed starting from 0. 
Nevertheless, we follow the convention to refer to the arrays indexes 
using the numbers of bodies in decay channel\footnote{
Elements with indexes equal to 0 are valid from C/C++ point of view,
but not used. Elements with indexes 1 are not used either: there are no
1-body decays.}.
The {\tt MAX\_DECAY\_MULTIPLICITY}
constant is defined in the {\tt src/Setup.H} source file, to be $20$ . 
In case you need 
to analyze more complex decay channels, you need to change this setting and 
recompile {\tt MC-Tester},
however we hope it is not very likely to happen. 

Example of use:

\texttt{// Assume that you need to analyze 5-body decays more thoroughly. }

\texttt{// In all 5-body decay channels, you are especially interested }

\texttt{// in analysis of histograms of mass of 3-body subsystems. }

\texttt{// Thus, you'd like to have the histograms more detailed: }

\texttt{Setup::nbins{[}5{]}{[}3{]}=256; }

\subsubsection{Setup::bin\_min}

Type:~ 2-D array: double{[}MAX\_DECAY\_MULTIPLICITY{]}{[}MAX\_DECAY\_MULTIPLICITY{]} 

Scope:~ generation 

Default: 0.0 

DESCRIPTION: Setup::bin\_min{[}n{]}{[}m{]} specifies the minimum bin value for
histogram of m-body invariant in the n-body decay mode.
Look at the example below and the description
of {\tt Setup::nbins} above for clarification. 

Example of use:

\texttt{// Assume that you need to analyze 5-body decays more thoroughly. }

\texttt{// In all 5-body decay channels, you are especially interested }

\texttt{// in analysis of histograms of mass of 3-body subsystems. }

\texttt{// You know that the mass of all subsystems will not be lower }

\texttt{// that 3.0GeV, and so should be the lower bound of histograms }

\texttt{Setup::bin\_min{[}5{]}{[}3{]}=3.0;}

\subsubsection{Setup::bin\_max}

Type:~ 2-D array: double{[}MAX\_DECAY\_MULTIPLICITY{]}{[}MAX\_DECAY\_MULTIPLICITY{]} 

Scope:~ generation

Default: 2.0 

DESCRIPTION: Setup::bin\_max{[}n{]}{[}m{]} specifies the maximum bin value 
for histogram of m-body invariant in the n-body decay mode.
 Look at the example below and the description
of {\tt Setup::nbins} above for clarification. 

Example of use:

\texttt{// Assume that you need to analyze 5-body decays more thoroughly. }

\texttt{// In all 5-body decay channels, you are especially interested }

\texttt{// in analysis of histograms of mass of 3-body subsystems. }

\texttt{// You know that the mass of all subsystems will not exceed }

\texttt{// 4.5GeV, and so should be the upper bound of histograms }

\texttt{Setup::bin\_max{[}5{]}{[}3{]}=4.5;}

\subsubsection{Setup::SetHistogramDefaults(int nbins, double min\_bin, double
max\_bin);}
\label{appendix.S.SetHistDefaults}

Type:~ function (static method of Setup class) 

Scope:~ generation 

DESCRIPTION: Sets up the default values for the number of bins, the minimum and maximum
bin for all the histograms. 

Note: {\em the dimensions and ranges of histograms processed at analysis step need to be
the same!}\footnote{The possibility to automatically re-bin the histograms will be
introduced in next release of {\tt MC-TESTER}.}

Example of use: 

\texttt{int default\_nbin=100; }

\texttt{double default\_min\_bin=0.0;}

\texttt{double default\_max\_bin=2.0;}

\texttt{Setup::SetHistogramDefaults(default\_nbin, default\_min\_bin, default\_max\_bin); }

\subsubsection{Setup::gen1\_path}

Type:~ char{*} 

Scope:~ analysis 

Default: {[}set at the generation step to the current directory{]} 

DESCRIPTION: This variable contains the path at which the first generator was
run, therefore indicates where the result file comes from. It is being initialized
at the generation step, however one may change it at the analysis step to any other
string. This path is printed at the first page of the booklet. It is also used
to search for a file named \char`\"{}version\char`\"{}. If it exists at the
path pointed by this variable, its contents are also printed in the booklet.
The version file is supposed to contain a short, one line description of 
a version of the code used by the generator, i.e. in {\tt TAUOLA} example it contains 
the strings \char`\"{}ALEPH\char`\"{}
or \char`\"{}CLEO\char`\"{} indicating two different branches of the generator 
code, which are being tested. 

Example of use:

\texttt{Setup::gen1\_path = \char`\"{}/my/new/path/of/first\_generator\char`\"{}}

\subsubsection{Setup::gen2\_path}

Type:~ char{*} 

Scope:~ analysis 

Default: {[}set at the generation step to the current directory{]}

DESCRIPTION: The same as above, for second generator. 

Example of use:

\texttt{Setup::gen2\_path = \char`\"{}/my/new/path/of/second\_generator\char`\"{}}

\subsubsection{Setup::user\_analysis}

Type:~ function pointer: double ({*})(TH1D{*},TH1D{*}) 

Scope:~ analysis 

Default: None - needs explicit specification in {\tt SETUP.C}. 

DESCRIPTION: Indicates a user-provided function to be used at the analysis step
to calculate 
%\char`\"{}shaping parameter\char`\"{}. 
{\tt SDP}.
We recommend to adopt the
choice of the function to your analysis.
Please refer to section \ref{sec:SDP-algorithms}.

Example of use:

\texttt{///// Setup analysis code (load it from file and set up)}

\texttt{gInterpreter->LoadMacro(\char`\"{}./MyAnalysis.C\char`\"{}); }

\texttt{Setup::user\_analysis=MyAnalysis; }

\texttt{printf(\char`\"{}Using Analysis code from file ./MyAnalysis.C \textbackslash{}n\char`\"{});}

\subsubsection{Setup::user\_event\_analysis}

Type:~ pointer to object: UserEventAnalysis{*} 

Scope:~ generation 

Default: None - functionality switched off

DESCRIPTION: Allows to specify the object (inheriting from
UserEventAnalysis class), which performs custom operations
on event record.

Please refer to files {\tt README.EVENT-ANALYSIS} and 
{\tt README-LC}.

Example of use:

\texttt{Setup::user\_event\_analysis=new LC\_EventAnalysis(); }

\subsubsection{Setup::SuppressDecay(int pdg);}
\label{option:SuppressDecay}

Type:~ function (static method of Setup class) 

Scope:~ generation 

DESCRIPTION: Suppresses decays of particles with PDG code given by the
parameter. The {\tt MC-TESTER} will treat these particles as if they were
stable.

Note: a maximum of 100 types of particles may be specified;

Example of use: 

\texttt{ Setup::SuppressDecay(111); // suppress pi0 decays }

\subsection{ {\tt F77} interface of {\tt MC-TESTER}.}
\label{appendix.F77}

A set of {\tt FORTRAN77} subroutines was provided to allow modification 
of some {\tt MC-TESTER} parameters. These functions are implemented in 
{\tt C++}, but can be called from the {\tt FORTRAN} program.

\subsubsection{\label{SETUP_in_F77}\texttt{SUBROUTINE MCSETUP( WHAT, VALUE) }}

\texttt{INTEGER WHAT }

\texttt{INTEGER VALUE}

{\it Description and parameters:}

\texttt{WHAT}: specifies what kind of value needs to be set 

\begin{itemize}
\item \texttt{WHAT=0} : Event record structure to be used:

\begin{itemize}
\item \texttt{VALUE=0} COMMON/HEPEVT/ in the 4k-D format
\item \texttt{VALUE=1} COMMON/LUJETS/ (i.e. Pythia 5.7)
\item \texttt{VALUE=2} COMMON/PYJETS/ (i.e. Pythia 6) 
\end{itemize}
\item \texttt{WHAT=1} : the generation stage: 

\begin{itemize}
\item \texttt{VALUE=1} or {\tt 2} for the first and the second generator, respectively.\\
Look at the {\tt TAUOLA} example -- the stage is introduced to two version of the code 
using a preprocessor.
\end{itemize}
\item \texttt{WHAT=2} : 
\begin{itemize}
\item 
\texttt{VALUE}= the PDG code of a particle, which decays 
we analyze.
\end{itemize}
\end{itemize}

\subsubsection{\texttt{SUBROUTINE MCSETUPHBINS(VALUE) }}

\texttt{INTEGER VALUE} 

{\it Description and parameters:}

Sets up the number of bins in histograms.

\subsubsection{\texttt{SUBROUTINE MCSETUPHMIN(VALUE) }}

\texttt{DOUBLE PRECISION VALUE} 

{\it Description and parameters:}

Sets up the value of the minimum bin in histograms.

\subsubsection{\texttt{SUBROUTINE MCSETUPHMAX(VALUE) }}

\texttt{DOUBLE PRECISION VALUE }

{\it Description and parameters:}

Sets up the value of the maximum bin in histograms.

\subsubsection{\texttt{SUBROUTINE MCSETUPHIST(NBODY,NHIST,NBINS,MINBIN,MAXBIN) }}

\texttt{INTEGER NBODY,NHIST,NBINS }

\texttt{DOUBLE PRECISION MINBIN,MAXBIN }

{\it Description and parameters:}

Sets up the parameters for histograms of {\tt NHIST}-body subsystems in 
{\tt NBODY}-bodies decay channel. 
{\tt NBINS} is the number of bins,
{\tt MINBIN} is the minimum bin value,
{\tt MAXBIN} is the maximum bin value.